\documentclass[preprint]{JHEP3} 

\JHEPspecialurl{http://jhep.sissa.it/JOURNAL/JHEP3.tar.gz}

\usepackage{epsfig,multicol,bbm}
\usepackage{axodraw}
\newcommand\fverbdo{\egroup\medskip\noindent%
            \fbox{\unhbox\fverbbox}\ }
\newcommand\fverbit{\egroup\item[\fbox{\unhbox\fverbbox}]}
\newbox\fverbbox

\title{Feynman rules for the rational part of the
Electroweak 1-loop amplitudes}

\author{}

\author{M.V. Garzelli\\
       Departamento de F\'{i}sica Te\'orica y del Cosmos y CAFPE
       Universidad de Granada, E-18071 Granada, Spain and \\
       INFN Milano, I-20133 Milano, Italy.
       E-mail: \email{garzelli@to.infn.it}}

\author{I. Malamos\\
        Department of Theoretical High Energy Physics,
        Institute for Mathematics, Astrophysics and Particle Physics, 
        Radboud Universiteit Nijmegen,  
        6525 AJ Nijmegen, the Netherlands. \\ 
        E-mail: \email{J.Malamos@science.ru.nl}}

\author{R. Pittau\\
       Departamento de F\'{i}sica Te\'orica y del Cosmos y CAFPE
       Universidad de Granada, E-18071 Granada, Spain.\\
       E-mail: \email{pittau@ugr.es}}

\abstract{
We present the complete set of Feynman rules producing 
the rational terms of kind ${\rm R_2}$
needed to perform any 1-loop calculation in the Electroweak Standard Model.
Our results are given both in the 't Hooft-Veltman and in the Four Dimensional 
Helicity regularization schemes.
We also verified, by using both the 't Hooft-Feynman gauge and
the Background Field Method, a huge set of Ward identities -up to 4-points- 
for the complete rational part of the Electroweak amplitudes.
This provides a stringent check of our results and, as a by-product, 
an explicit test of the gauge invariance of the Four Dimensional Helicity 
regularization scheme in the complete Standard Model at 1-loop.  
The formulae presented in this paper provide the last missing piece for
completely automatizing, in the framework of the OPP method, the 1-loop 
calculations in the $SU(3) \times SU(2) \times U(1)$ Standard Model.}


\begin{document}

\newcounter{im}
\setcounter{im}{0}
\newcommand{\exampleSp}{\stepcounter{im}\includegraphics[scale=0.9]{SpinorExamples_\arabic{im}.eps}}
\newcommand{\myindex}[1]{\label{com:#1}\index{{\tt #1} & pageref{com:#1}}}
\renewcommand{\topfraction}{1.0}
\renewcommand{\bottomfraction}{1.0}
\renewcommand{\textfraction}{0.0}
\newcommand{\nn}{\nonumber \\}
\newcommand{\eqn}[1]{eq.~\ref{eq:#1}}
\newcommand{\be}{\begin{equation}}
\newcommand{\ee}{\end{equation}}
\newcommand{\ba}{\begin{array}}
\newcommand{\ea}{\end{array}}
\newcommand{\bea}{\begin{eqnarray}}
\newcommand{\eea}{\end{eqnarray}}
\newcommand{\bqa}{\begin{eqnarray}}
\newcommand{\eqa}{\end{eqnarray}}
\newcommand{\nl}{\nonumber \\}
\def\db#1{\bar D_{#1}}
\def\zb#1{\bar Z_{#1}}
\def\d#1{D_{#1}}
\def\tld#1{\tilde {#1}}
\def\slh#1{\rlap / {#1}}
\def\eqn#1{eq.~(\ref{#1})}
\def\eqns#1#2{Eqs.~(\ref{#1}) and~(\ref{#2})}
\def\eqnss#1#2{Eqs.~(\ref{#1})-(\ref{#2})}
\def\fig#1{Fig.~{\ref{#1}}}
\def\figs#1#2{Figs.~\ref{#1} and~\ref{#2}}
\def\sec#1{Section~{\ref{#1}}}
\def\app#1{Appendix~\ref{#1}}
\def\tab#1{Table~\ref{#1}}
\def\cg{c_\Gamma}
\newcommand{\bfig}{\begin{center}\begin{picture}}
\newcommand{\efig}[1]{\end{picture}\\{\small #1}\end{center}}
\newcommand{\flin}[2]{\ArrowLine(#1)(#2)}
\newcommand{\ghlin}[2]{\DashArrowLine(#1)(#2){5}}
\newcommand{\wlin}[2]{\DashLine(#1)(#2){2.5}}
\newcommand{\zlin}[2]{\DashLine(#1)(#2){5}}
\newcommand{\glin}[3]{\Photon(#1)(#2){2}{#3}}
\newcommand{\gluon}[3]{\Gluon(#1)(#2){5}{#3}}
\newcommand{\lin}[2]{\Line(#1)(#2)}
\newcommand{\sof}{\SetOffset}



\section{Introduction}
The complete automation of the 1-loop calculations for LHC and ILC physics
is nowadays a feasible task~\cite{nlopapers}. 
The advent of the OPP reduction method~\cite{opp},
together with the concept of multiple cuts~\cite{britto}, 
allowed to revitalize the 
Unitarity Techniques~\cite{unitarity}, by reducing the computation of
1-loop amplitudes to a problem with the same conceptual 
complexity of a tree level calculation, resulting in achievements that 
were inconceivable only a few years ago~\cite{proof}.

The main idea of the OPP based techniques is directly extracting, 
from the 1-loop amplitude, 
the coefficients of the (known) scalar loop functions. This task can 
be reached in a completely numerical way 
by  {\em opening the loop} and transforming the 1-loop amplitude 
in a tree level object with 2 more legs, that can be calculated, 
at the {\em integrand level}, by using the same
recursion relations~\cite{recursions} that allow a very efficient
computation of complicated multi-leg tree 
level processes~\cite{alpgen,helac}. 
A second possible 
option is that one of the so-called Generalized Unitarity 
methods~\cite{genunit}, 
where tree-level amplitudes are {\em glued together}.

Both procedures only allow the extraction of 
the Cut Constructible (CC) part of the amplitude in 4 dimensions, while a 
left over piece, the rational part ${\rm R}$, needs to be derived separately.
In the Generalized Unitarity approaches, that is achieved by computing
the amplitude in different numbers of space-time dimensions, or 
via bootstrapping techniques~\cite{boot}, while, 
in the OPP approach, ${\rm R}$ is split in 2 pieces ${\rm R= R_1+R_2}$. The first 
piece, ${\rm R_1}$, is derivable in the same framework used to reconstruct the 
CC part of the amplitude, while ${\rm R_2}$ is computable 
through a special set Feynman rules for the theory at hand~\cite{rational}, 
to be used in a tree level-like computation. 

Such a set of ${\rm R_2}$ Feynman rules has been already derived for QED 
in~\cite{rational} and for QCD in~\cite{qcdrational}, and it is the main aim of the present paper 
to present the complete set of the ${\rm R_2}$ Feynman rules for the Standard Model (SM)
of the Electroweak (EW) interactions. 
In addition, as a by-product, we use the derived formulae to explicitly check 
the gauge invariance of the Four Dimensional Helicity regularization 
scheme in the EW sector at 1-loop, the motivation being that this is a very 
well studied subject in QCD~\cite{fdhqcd}, but, in our knowledge, very little can be found in the literature for the full EW Standard Model.

The outline of the paper is as follows. In section~\ref{sec:2} we remind some
facts on the origin of ${\rm R}$ and on the splitting ${\rm R}= {\rm R_1}+{\rm R_2}$. Section~\ref{sec:3} contains the complete
list of all possible special ${\rm R_2}$ EW SM vertices 
in the 't Hooft-Feynman gauge and, in section~\ref{sec:4}, 
we describe the tests we performed on our formulae and our findings.
Finally, our conclusions are drawn in section~\ref{sec:5} and, in the appendix, 
we collect a list of Ward identities.

\section{Theory, facts and conjectures on ${\rm R}$, ${\rm R_1}$ and ${\rm R_2}$  \label{sec:2}}
Before carrying out our program, we spend a few words on
the origin of ${\rm R}$. Our starting point is the general expression for the
{\it integrand} of a generic $m$-point
one-loop (sub-) amplitude
\bqa
\label{eq:1}
\bar A(\bar q)= \frac{\bar N(\bar q)}{\db{0}\db{1}\cdots \db{m-1}}\,,~~~
\db{i} = ({\bar q} + p_i)^2-m_i^2\,,
\eqa
where ${\bar q}$ is the integration momentum and where dimensional 
regularization is assumed, so that
a bar denotes objects living
in $n=~4+\epsilon$ dimensions and a tilde represents $\epsilon$-dimensional
quantities.
When a $n$-dimensional index is contracted with a 4-dimensional vector
$v_\mu$, the $4$-dimensional part is automatically selected. 
For example
\bqa
\label{noeps}
\bar q \cdot v &\equiv& (q+ {\tld q}) \cdot v\,= q \cdot v\,,~~~
\rlap/ {\bar v} \equiv  \bar \gamma_{\bar \mu}\, v^\mu = \rlap /v\,
~~~{\rm and}~~~ {\bar q}^2 = q^2 + {\tld q}^2\,.
\eqa
The numerator function $\bar{N}(\bar q)$ can be 
split into a $4$-dimensional plus an $\epsilon$-dimensional part
\bqa
\label{eq:split}
\bar{N}(\bar q) = N(q) + \tld{N}(\tld{q}^2,q,\epsilon)\,.
\eqa
$N(q)$ lives in $4$ dimensions, and can be therefore expanded in terms of 
$4$-dimensional denominators
\bqa
\d{i} = ({q} + p_i)^2-m_i^2 = \db{i} - {\tld q}^2\,.
\eqa
Some among the coefficients in this expansion are interpreted, in the OPP 
method, as the desired coefficients of the 1-loop scalar integrals and 
can be determined numerically, while
the mismatch between this expansion in terms of $4$-dimensional denominators, 
and the $n$-dimensional denominators appearing in eq.~\ref{eq:1}, 
is the origin of the rational terms ${\rm R_1}$. There exist at least 
two ways~\cite{sixphoton,cuttools} to compute ${\rm R_1}$, 
which allow to determine it by means of a purely numerical knowledge 
of the 4-dimensional CC part of the amplitude, while this does not seem to 
be possible
for ${\rm R_2}$, whose origin is the term
$\tld{N}(\tld{q}^2,q,\epsilon)$ in eq.~\ref{eq:split}, after 
integration over the loop momentum:
\bqa
\label{eqr2}
{\rm R_2} \equiv  \frac{1}{(2 \pi)^4}\int d^n\,\bar q
\frac{\tld{N}(\tld{q}^2,q,\epsilon)}{\db{0}\db{1}\cdots \db{m-1}} \,.
\eqa
However, ${\rm R_2}$ can be computed by extracting 
$\tld{N}(\tld{q}^2,q,\epsilon)$  from any given {{\em integrand} 
$\bar A (\bar q)$, which can be achieved 
by splitting, in the  analytic expression of the numerator function, 
the $n$-dimensional 
integration momentum
${\bar q}$, the $n$-dimensional gamma matrices
$\bar \gamma_{\bar \mu}$  and the $n$-dimensional metric tensor
$\bar g^{\bar \mu \bar \nu}$ into a $4$-dimensional
component plus remaining pieces:
\bqa
\label{qandg}
\bar q                 &=& q + \tld{q}\,, \nl
\bar \gamma_{\bar \mu} &=&  \gamma_{\mu}+ \tld{\gamma}_{\tld{\mu}}\,,\nl
 \bar g^{\bar \mu \bar \nu}  &=&  g^{\mu \nu}+  \tld{g}^{\tld{\mu} \tld{\nu}}\,.
\eqa
Therefore, a practical way to determine ${\rm R_2}$ is computing 
analytically, by means of Feynman diagrams, once for all
and with the help of eq.~\ref{qandg}, tree-level like Feynman rules, 
namely effective vertices, by calculating the ${\rm R_2}$ part coming from
all possible one-particle irreducible Green functions of the theory at hand, 
up to four external legs.
The fact that four external legs are enough to account for ${\rm R_2}$ 
is guaranteed by the ultraviolet nature of the rational terms, 
proved in~\cite{directcomp1}.
This property does not hold, instead, for ${\rm R_1}$, that, diagram 
by diagram, can give non vanishing contributions to any  
one-particle irreducible $m$-point function, because, even when finite, the 
tensor integrals generating ${\rm R_1}$ are eventually expressed, via tensor 
reduction, in terms of linear combinations of 1-loop scalar functions that can 
be ultraviolet divergent. This fact prevents the possibility
of calculating a finite set of effective vertices reproducing ${\rm R_1}$.

Eq.~\ref{eqr2} generates a set of simple basic integrals with up to 4 denominators, containing powers of $\tld{q}$ and $\epsilon$ in the numerator. 
A list that exhausts all possibilities in the $\xi = 1$ 't Hooft-Feynman gauge 
can be found in~\cite{qcdrational}.
Notice, however, that, according to the chosen regularization scheme,
results may differ. In eq.~\ref{eqr2} we assume the
't Hooft-Veltman (HV) scheme, while in the Four Dimensional Helicity scheme
(FDH), any explicit $\epsilon$ dependence in the numerator function
is discarded before integration, such that
\bqa
\label{eqr2fdh}
{\rm R_2} \Bigl |_{FDH} =  \frac{1}{(2 \pi)^4}\int d^n\,\bar q
\frac{\tld{N}(\tld{q}^2,q,\epsilon= 0)}{\db{0}\db{1}\cdots \db{m-1}} \,.
\eqa

The asymmetric role played by ${\rm R_1}$ and ${\rm R_2}$ 
is somewhat annoying. As we have seen, ${\rm R_1}$ is directly 
connected with the
($4$-dimensional) CC part of the amplitude, and can be computed, even numerically, without any reference to Feynman diagrams, while 
${\rm R_2}$ requires an analytic determination in terms of Feynman diagrams, 
so that one would like 
to be able to put both pieces on the same footing. 
Unfortunately, no easy direct 
connection between ${\rm R_2}$ and the CC part of the amplitude has been found so far (at least within our treatment at the {\em integrand level}) and,
in the rest of this paragraph, we speculate a bit on this subject.
 
Reconstructing ${\rm R_2}$  numerically would require to detect ``signs'' of it in the CC part.
For example, one could naively think that, by
looking at any $q^2$ in the CC part, the $\tld{q}^2$ dependence
could be inferred via the replacement 
\bqa
\label{eq:repl}
q^2 \to q^2 + \tld{q}^2\,.
\eqa
However, such a dependence is impossible to reconstruct numerically, when remaining in 4 dimensions, as it can be illustrated by considering the 
following simple 3-point sub-amplitude:
\bqa
\label{eqex}
{\rm A  } \equiv  \frac{1}{(2 \pi)^4}\int d^n\,\bar q
\frac{(q \cdot \ell_3)(q \cdot \ell_4) }{\db{0}\db{1}\db{2}} \,,
\eqa
where
\bqa
\ell_3^\mu = <\ell_1| \gamma^\mu | \ell_2]\,,~~
\ell_4^\mu = <\ell_2| \gamma^\mu | \ell_1]\,
~~{\rm with}~~
\ell_{1,2}^2= 0\,.
\eqa
From the one hand, the $4$-dimensional numerator $(q \cdot \ell_3)(q \cdot \ell_4)$ in eq.~\ref{eqex} does not 
contain any $q^2$ to be continued through the replacement of 
eq.~\ref{eq:repl}. On the other hand, it can be 
manipulated as follows
\bqa
(q \cdot \ell_3)(q \cdot \ell_4)= 4\,(q \cdot \ell_1)(q \cdot \ell_2)-
2\,q^2\,(\ell_1 \cdot \ell_2)\,,
\eqa
and now the shift of eq.~\ref{eq:repl} would produce a $\tld{q}^2$ contribution, 
in disagreement with our previous finding. We therefore conclude that
not enough information is present in the 4-dimensional part to reconstruct
${\rm R_2}$. This is the reason why one is forced to work analytically  
in $n$ dimensions to reconstruct the ${\rm R_2}$ contribution
\footnote{
In other approaches~\cite{genunit}, a numerical determination of the
whole ${\rm R}$ contribution can be achieved, but at the price of 
explicitly computing numerically the amplitude in 4, 6 and 8 dimensions.}. 

Nevertheless, based on a simple reasoning, one argues
that some gauge invariance properties of the $4$-dimensional
part of the amplitude should be transferred to ${\rm R_2}$.
In fact, for physical processes, the sum of ${\rm R_1} + {\rm R_2}$ is 
gauge invariant. On the other hand, ${\rm R_1}$
can be fully reconstructed from the $4$-dimensional, gauge invariant,
CC part of the amplitude, meaning that, by changing gauge, 
the same expressions for ${\rm R_1}$ should be found, and, as a consequence, 
also {\em the same result for ${\rm R_2}$}. 
This should be off course only true for amplitudes with
physical external particles, because different gauges may have, in general, 
a different content in terms unphysical external fields. Therefore 
one can conjecture that
\begin{center}
{\em The ${\rm R_2}$ part of a physical amplitude gives the same result when computed in any gauge \footnote{This does not mean that the ${\rm R_2}$ part
of the Green functions satisfy the Ward identities 
separately from ${\rm R_1}$, as we have checked explicitly.}
.} 
\end{center}
This conjecture, being rather strong, should be proved with an actual 
calculation. Unfortunately, such a calculation would require to extend the 
set of basic integrals in~\cite{qcdrational} to be able to deal with 
non-renormalizable gauges. That is beyond the scope of this work, and 
we leave it for a future publication. 

In the present paper, we fix the gauge to be the  
the 't Hooft-Feynman one and we derive all of the effective Electroweak
${\rm R_2}$ Feynman rules by applying the splittings of eq.~\ref{qandg} 
Feynman diagram by Feynman diagram. For the interested reader, 
explicit examples of this technique can be found in ~\cite{qcdrational}.

\section{Results \label{sec:3}}
In this section, we give the complete list of the
effective Electroweak vertices contributing to ${\rm R_2}$ in
the 't Hooft-Feynman gauge
\footnote{They can be also found as a {\tt FORM}~\cite{form} output in
http://www.ugr.es/local/pittau/CutTools.}. 
A parameter $\lambda_{HV}$ is introduced in our formulae such that 
$\lambda_{HV}= 1$ corresponds to the 't Hooft-Veltman scheme
and $\lambda_{HV}= 0$ to the FDH scheme of eq.~\ref{eqr2fdh}.
We used the Feynman rules given in~\cite{denner} and
our notations are
as follows: $l_1= e$, $l_2= \mu$, $l_3= \tau$, $l_4= \nu_e$, $l_5= \nu_\mu$,
$l_6= \nu_\tau$ and $q_1= d$, $q_2= u$, $q_3= s$, $q_4= c$, $q_5= b$, $q_6= t$.
In addition, 
$e_1= e$, 
$e_2= \mu$, 
$e_3= \tau$,
$\nu_1= \nu_e$, 
$\nu_2= \nu_\mu$,
$\nu_3= \nu_\tau$ and 
$u_1= u$, $u_2= c$, $u_3= t$, $d_1= d$, $d_2= s$, $d_3= b$.
When appearing as external particles, $l$, $\nu_l$, $u$ and $d$  
stand for the three charged leptons, the three (massless) neutrinos, 
the three up-type quarks and  
the three down-type quarks, respectively. 
Effective vertices with external quarks are always 
understood to be diagonal  in the color space. 
Finally, $N_{col}$ is the number 
of colors and $V_{u_i d_j}$ are CKM matrix elements.
Occasionally, combinations such as 
\bqa
\sum_{i,j=1}^{3}  
\left(
V_{u_i d_j} V_{d_j u_i}^\dagger 
\right) = 3\,~~~{\rm and}~~~ \sum_{i=1}^{3} 1 = 3  
\nonumber
\eqa
appear in our formulae. In such cases, we do not explicitly work out
the sum in order to make our results also readable family by family.

A last comment is in order with respect to our 
treatment of $\gamma_5$ in vertices 
containing fermionic lines. When computing all contributing Feynman diagrams, 
we pick up a ``special'' vertex  in the loop and anticommute all $\gamma_5$'s 
to reach it before performing the $n$-dimensional algebra, and,  when a trace 
is present, we start reading it from this vertex. This treatment
produces, in general, a term proportional to the totally antisymmetric $\epsilon$ 
tensor, whose coefficient may be different 
depending on the choice of the ``special'' 
vertex. However, when summing over all quantum numbers of each fermionic 
family, we checked that all contributions proportional to $\epsilon$ cancel. 
In addition, we explicitly verified that our results satisfy
the large set of Ward identities given in appendix~\ref{appa}.
\subsection{\underline{Electroweak effective vertices with 2 external legs}}
In this section, we give the complete list of the non vanishing
2-point ${\rm R_2}$ effective vertices.
\subsubsection{{Scalar-Scalar effective vertices}}
The generic effective vertex is

\vspace{0.3cm}

\begin{center}
\fbox{
  \begin{picture}(200,50)
   \SetScale{0.5}
    \SetWidth{0.5}
    \SetColor{Black}
    \SetOffset(0,55)
    \Vertex(142,-62){6.0}
    \DashLine(56,-62)(142,-62){8}
    \DashLine(142,-62)(228,-62){8}
    \Text(124,-31)[]{$S_2$} 
    \Text(18,-31)[]{$S_1$} 
    \Text(138,-31)[l]{$\displaystyle = \frac{ie^2}{16 \pi^2 s_w^2} C$}
  \end{picture}
}
\end{center}
with the actual values of $S_1$, $S_2$ and $C$
\bqa
%
%
H\chi~~:~~C & = & 0 
\nonumber \\ 
\nonumber \\ 
%
%
HH~~:~~C & = & 
\frac{m_\phi^2}{4} + \frac{m_\chi^2}{8 c_w^2} 
+\frac{1-12\lambda_{HV}}{4}\left(1+\frac{1}{2c_w^4}\right) m_W^2  
- \left(1+\frac{1}{2 c_w^2} \right) \frac{p^2}{12} + K
\nonumber \\
\nonumber \\
%
%
\chi\chi~~:~~C & = & 
\frac{m_\phi^2}{4} + \frac{m_H^2}{8 c_w^2} 
+\frac{1-4\lambda_{HV}}{4} \left(1+\frac{1}{2c_w^4}\right)  m_W^2 -
\left(1+\frac{1}{2 c_w^2} \right) \frac{p^2}{12} + K
\nonumber \\
\nonumber \\
%
%
\phi^-\phi^+ ~~:~~C & = &
\frac{m_H^2+m_\chi^2}{8} 
+ \frac{\left(3-4\lambda_{HV}\right) c_w^4 -2 c_w^2  + \left(\frac{1}{2}-2\lambda_{HV}\right)}{c_w^4}\frac{m_W^2}{4}
+ \frac{m_\phi^2}{8 c_w^2} 
\nonumber \\
&-& \left(1 + \frac{1}{2 c_w^2} \right) 
\frac{p^2}{12} 
 +  
\frac{1}{2 m_W^2 } \left[
\sum_{i= 1}^{3}
\left(m_{e_i}^2
\left(m_{e_i}^2-\frac{p^2}{3}  \right)\right)  
\right. \nonumber \\ 
& + & \left. N_{\mathrm{col}} \sum_{i,j=1}^{3}  
\left(
V_{u_i d_j} V_{d_j u_i}^\dagger
\left(m_{u_i}^2 + m_{d_j}^2\right) 
\left(m_{u_i}^2 + m_{d_j}^2-\frac{p^2}{3}\right)  
\right)
\right]
\eqa
where
\bqa
K &=& 
\frac{1}{m_W^2}\left[
\sum_{i=1}^{6} \left( m_{l_i}^2\left(m_{l_i}^2 - \frac{p^2}{6}
\right)\right)
+ N_{\mathrm{col}}
\sum_{i=1}^{6} \left( m_{q_i}^2\left(m_{q_i}^2 - \frac{p^2}{6}
\right)
\right)
\right] 
\eqa
\subsubsection{{Vector-Vector effective vertices}}
The generic effective vertex is

\vspace{0.3cm}

\begin{center}
\fbox{
  \begin{picture}(250,50) 
    \SetOffset(0,42)
    \SetScale{0.5}
    \SetWidth{0.5}
    \SetColor{Black}
    \Photon(65,-34)(133,-34){5.5}{4}
    \Vertex(134,-34){6.0}
    \Photon(135,-34)(199,-34){5.5}{4}
    \Text(50,-3)[]{{\Black{$p$}}}
    \LongArrow(75,-20)(110,-20)  
    \Text(21,-17)[]{$V_{1\alpha}$}
    \Text(113,-17)[]{$V_{2\beta}$}
    \Text(130,-14)[l]{$\displaystyle
= \frac{ie^2}{\pi^2} 
\left(C_1 p_\alpha p_\beta + C_2 g_{\alpha \beta}\right)
$}
  \end{picture}
}
\end{center}
with the actual values of $V_1$, $V_2$, $C_1$ and $C_2$
\bqa
%
%
AA~~:~~C_1 & = & - \frac{1}{24} \lambda_{HV} \nonumber \\
C_2 & = & \frac{1}{8}\left[p^2\left(
\frac{1}{6}+\frac{\lambda_{HV}}{3}\right) - m_W^2 \right] 
 -  \frac{1}{4} \left[
\sum_{i=1}^{6}\left(Q_{l_i}^2
\left(m_{l_i}^2 - \frac{1}{6} p^2\right)\right)
\right. \nonumber \\
& + &  \left. 
N_{\mathrm{col}}\sum_{i=1}^{6}\left(Q_{q_i}^2
\left(m_{q_i}^2 - \frac{1}{6} p^2\right)\right)
\right] 
\nonumber \\
\nonumber \\
%
%
AZ~~:~~C_1 & = & \frac{1}{24} \frac{c_w}{s_w} \lambda_{HV}\, \nonumber \\
C_2 & = & - \frac{1}{8}\frac{c_w}{s_w}\left[p^2\left( 
\frac{1}{6}+\frac{\lambda_{HV}}{3}\right) - m_W^2 \right] 
 +  \frac{1}{4 c_w} \left[ \sum_{i=1}^{6}\left(
\left(\frac{Q_{l_i} I_{3l_i}}{2 s_w} - Q_{l_i}^2 s_w\right)
\right. \right.
\nonumber \\
&\times& \left. \left.
\left(m_{l_i}^2 - \frac{1}{6} p^2\right)\right)
+ N_{\mathrm{col}} \sum_{i=1}^{6}\left(
\left(\frac{Q_{q_i} I_{3q_i}}{2 s_w} - Q_{q_i}^2 s_w\right)
\left(m_{q_i}^2 - \frac{1}{6} p^2\right)\right)
\right]
\nonumber \\
\nonumber \\
%
%
ZZ~~:~~C_1 & = & - \frac{1}{24} \frac{c_w^2}{s_w^2} \lambda_{HV}\nonumber \\
C_2 & = & \frac{1}{8}\frac{c_w^2}{s_w^2}\left[p^2\left(\frac{1}{6}+\frac{\lambda_{HV}}{3}\right)- m_W^2 \right] +  
\frac{1}{4 c_w^2} 
\left[
\sum_{i=1}^{6}\left(
\left(Q_{l_i} I_{3l_i} - \frac{I_{3l_i}^2}{2 s_w^2} - Q_{l_i}^2 s_w^2\right)
\right. \right.
\nonumber \\
&\times& \left. \left.
\left(m_{l_i}^2 - \frac{1}{6} p^2\right)\right)
+N_{\mathrm{col}}\sum_{i=1}^{6}\left(
\left(Q_{q_i} I_{3q_i} - \frac{I_{3q_i}^2}{2 s_w^2} - Q_{q_i}^2 s_w^2\right)
\left(m_{q_i}^2 - \frac{1}{6} p^2\right)\right)
\right]
\nonumber \\
\nonumber \\
%
%
W^-W^+~~:~~C_1 & = & - \frac{1}{24 s_w^2} \lambda_{HV}\nonumber \\
C_2 & = & \frac{1}{8 s_w^2}\left[p^2\left(\frac{1}{6}+\frac{\lambda_{HV}}{3}\right) - m_W^2 \right]
- \frac{1}{32 s_w^2} 
\left[ \sum_{i=1}^3 \left(m_{e_i}^2 - \frac{p^2}{3}\right) 
\right.
\nonumber \\
& + &  \left. N_{\mathrm{col}}\sum_{i,j=1}^{3}
\left(
V_{u_i d_j} V_{d_j u_i}^\dagger 
\left(m_{u_i}^2 + m_{d_j}^2 - \frac{p^2}{3}\right) 
\right)
\right]
\eqa

\subsubsection{{Fermion-Fermion effective vertices}}
The generic effective vertex is

\vspace{0.3cm}

\begin{center}
\fbox{
  \begin{picture}(320,50) 
   \SetOffset(0,55)
    \SetWidth{0.5}
    \SetScale{0.5}
    \SetColor{Black}
    \ArrowLine(62,-62)(138,-62)
    \ArrowLine(141,-62)(221,-62)
    \Vertex(142,-63){6.0}
    \Text(50,-18)[]{{\Black{$p$}}}
    \LongArrow(75,-50)(110,-50)  
    \Text(22,-31)[]{$F_1$}
    \Text(123,-31)[]{$\bar F_2$}
    \Text(135,-31)[l]{$\displaystyle = \frac{ie^2}{\pi^2}
\left[\left(C_- \Omega^- + C_+\Omega^+\right)\rlap/p + C_0
\right]\lambda_{HV}
$}
  \end{picture}
}
\end{center}
with the actual values of $F_1$, $\bar F_2$, $C_-$, $C_+$ and $C_0$
\bqa
%
%
u \bar u ~~:~~C_- & = & \frac{1}{16} \frac{Q_u^2}{c_w^2} \nonumber\\ 
C_+ & = & \frac{1}{16} \left(\frac{I_{3u}^2}{s_w^2 c_w^2}
- \frac{ 2 Q_{u} I_{3u}}{c_w^2} + \frac{Q_u^2}{c_w^2}
+ \frac{1}{2 s_w^2} 
\sum_{j=1}^3 \left(V_{u d_j} V_{d_j u}^\dagger\right)
 \right) \nonumber \\
C_0 & = & \frac{m_{u} Q_{u} }{8 c_w^2} \left( Q_{u} - I_{3u} \right) 
\nonumber \\
\nonumber \\
%
%
d \bar d ~~:~~C_- & = & \frac{1}{16} \frac{Q_{d}^2}{c_w^2} \nonumber \\
C_+ & = & \frac{1}{16} \left(
\frac{I_{3{d}}^2}{s_w^2 c_w^2}
- \frac{ 2 Q_{d} I_{3d}}{c_w^2} + \frac{Q_{d}^2}{c_w^2}
+ \frac{1}{2 s_w^2}
\sum_{i=1}^3 \left(V_{u_i d} V_{d u_i}^\dagger\right) 
\right) \nonumber \\
C_0 & = & \frac{m_{d} Q_{d}}{8 c_w^2} \left( Q_{d} - I_{3d}  \right) 
\nonumber \\
\nonumber \\
%
%
l \bar l ~~:~~C_- & = &  \frac{1}{16} \frac{Q_{l}^2}{c_w^2}  \nonumber \\
C_+ & = &  \frac{1}{16} \left(\frac{I_{3l}^2}{s_w^2 c_w^2}
- \frac{ 2 Q_{l} I_{3l}}{c_w^2} + \frac{Q_{l}^2}{c_w^2}
+ \frac{1}{2 s_w^2} \right) \nonumber \\
C_0 & = & \frac{m_{l}Q_{l}}{8 c_w^2} \left(  Q_{l} - I_{3l} \right) 
\nonumber \\
\nonumber \\
%
%
\nu_l \bar \nu_l ~~:~~C_- & = &  0 \nonumber \\
C_+ & = &  \frac{1}{32 s_w^2} \left(\frac{1}{2 c_w^2}
+ 1 \right) \nonumber \\
C_0 & = & 0
\eqa
%
%
\subsection{\underline{Electroweak effective vertices with 3 external legs}}
We list here the 3-point ${\rm R_2}$ effective vertices.
\subsubsection{{Scalar-Fermion-Fermion effective vertices}}
The generic effective vertex is

\vspace{0.3cm}

\begin{center}
\fbox{
  \begin{picture}(210,65) 
   \SetOffset(-20,25)
    \SetWidth{0.5}
    \SetScale{0.5}
    \SetColor{Black}
    \Vertex(146,15){6.0}
    \DashLine(72,15)(144,15){8}
    \ArrowLine(200,57)(146,15)
    \ArrowLine(146,15)(200,-27)
\Text(33, 7.5)[r]{$S$}
\Text(102, 30)[l]{$F_1$}
\Text(102,-15)[l]{$\bar F_2$}
\Text(115,7.5)[l]{$\displaystyle = \frac{e^3}{\pi^2} 
(C_{-}\Omega^{-} + C_{+}\Omega^{+})$}
  \end{picture}
}
\end{center}
with the actual values of $S$, $F_1$, $\bar F_2$, $C_-$ and $C_+$
\bqa
%
%
H u \bar u ~~:~~C_- &=& \frac{i m_u}{8 m_W s_w} \left[
\frac{\left(1+\lambda_{HV} \right)Q_u^2}{2c_w^2}
+\frac{1}{16 s_w^2}\sum_{j=1}^3
\left(
V_{u d_j}V_{d_ju}^\dagger
\right)
+\frac{I_{3u}}{c_w^2}
\left(
\frac{I_{3u}}{8 s_w^2} 
\right. \right. \nonumber \\
&-& \left. \left. \frac{\left(1+\lambda_{HV} \right)Q_u}{2}
\right) 
+\frac{1}{16 m_W^2 s_w^2}
\sum_{j=1}^3 
\left(m_{d_j}^2 V_{ud_j}V_{d_ju}^\dagger\right) 
\right] 
\nonumber \\
C_+ &=& C_-  
\nonumber \\
\nonumber \\
%
%
H d \bar d ~~:~~C_- &=& \frac{i m_d}{8 m_W s_w} \left[
\frac{ \left(1+\lambda_{HV} \right)Q_d^2}{2c_w^2}
+\frac{1}{16 s_w^2}\sum_{i=1}^3
\left(
V_{u_i d}V_{d u_i}^\dagger
\right)
+\frac{I_{3d}}{c_w^2}
\left(
\frac{I_{3d}}{8 s_w^2} 
\right. \right. \nonumber \\
&-&  \left. \left. \frac{\left(1+\lambda_{HV} \right)Q_d}{2}
\right) 
+\frac{1}{16 m_W^2 s_w^2}
\sum_{i=1}^3 
\left(m_{u_i}^2 V_{u_id}V_{du_i}^\dagger\right) 
\right] 
\nonumber \\
C_+ &=& C_-  
\nonumber \\
\nonumber \\
%
%
H l \bar l ~~:~~C_- &=& \frac{i m_l}{8 m_W s_w} \left[
\frac{ \left(1+\lambda_{HV} \right)Q_l^2}{2c_w^2}
+\frac{1}{16 s_w^2}
+\frac{I_{3l}}{c_w^2}
\left( \frac{I_{3l}}{8 s_w^2} - \frac{\left(1+\lambda_{HV} 
\right)Q_l}{2} \right) 
\right] 
\nonumber \\
C_+ &=& C_-  
\nonumber \\
\nonumber \\
%
%
H \nu_l \bar \nu_l ~~:~~C_- &=& 0
\nonumber \\
C_+ &=& 0 
\nonumber \\
\nonumber \\
%
%
\chi u \bar u ~~:~~C_- & = & -\frac{m_u}{4 m_W s_w} \left[
\frac{ \left(1+\lambda_{HV} \right)Q_u^2 I_{3u}}{2c_w^2}
+\frac{1}{32 s_w^2}\sum_{j=1}^3
\left(
V_{u d_j}V_{d_ju}^\dagger
\right)
+\frac{I_{3u}}{c_w^2}
\left(
\frac{1}{32 s_w^2} 
\right. \right. \nonumber \\
&-& \left. \left.  \frac{\left(1+\lambda_{HV} \right)Q_u I_{3u}}{2}
\right) 
- \frac{1}{16 m_W^2 s_w^2}
\sum_{j=1}^3 
\left(m_{d_j}^2 I_{3d_j} V_{ud_j}V_{d_ju}^\dagger\right) 
\right]
\nonumber \\
C_+ &=& -C_-  
\nonumber \\
\nonumber \\
%
%
\chi d \bar d ~~:~~C_- & = & -\frac{m_d}{4 m_W s_w} \left[
\frac{\left(1+\lambda_{HV} \right)Q_d^2 I_{3d}}{2c_w^2}
-\frac{1}{32 s_w^2}\sum_{i=1}^3
\left(
V_{u_i d}V_{d u_i}^\dagger
\right)
+\frac{I_{3d}}{c_w^2}
\left(
\frac{1}{32 s_w^2} 
\right. \right. \nonumber \\
&-& \left. \left. \frac{\left(1+\lambda_{HV} \right)Q_d I_{3d}}{2}
\right) 
-\frac{1}{16 m_W^2 s_w^2}
\sum_{i=1}^3 
\left(m_{u_i}^2 I_{3u_i} V_{u_id}V_{d u_i}^\dagger\right) 
\right]
\nonumber \\
C_+ &=& -C_-  
\nonumber \\
\nonumber \\
%
%
\chi l \bar l ~~:~~C_- & = & -\frac{m_l}{4 m_W s_w} \left[
\frac{\left(1+\lambda_{HV} \right)Q_l^2 I_{3l}}{2c_w^2}
-\frac{1}{32 s_w^2}
+\frac{I_{3l}}{c_w^2}
\left(
\frac{1}{32 s_w^2} - \frac{\left(1+\lambda_{HV} \right)Q_l I_{3l}}{2}
\right) \right.\nonumber\\
&-& \left. \frac{m_l^2 I_{3l}}{8 m_W^2 s_w^2}
\left( 
-\frac{1}{4} + I_{3l}^2
\right)
\right]
\nonumber\\
C_+ &=& -C_-  
\nonumber \\
\nonumber \\
%
%
\chi \nu_l \bar \nu_l ~~:~~C_- & = & 0
\nonumber\\
C_+ &=& 0
\nonumber \\
\nonumber\\
%
%
\phi^- u \bar{d} ~~:~~
C_- & = & -\frac{im_d V_{du}^\dagger}{4 \sqrt{2} m_W s_w} \left[
\frac{1}{c_w^2}
\left(
\frac{-1}{16} - \frac{\left(1+\lambda_{HV} \right)Q_u Q_d}{2}
\right)
- \frac{3}{32 s_w^2}
\right. \nonumber\\
&-&\left.
\frac{m_u^2}{16 s_w^2 m_W^2}
+\frac{I_{3u}}{c_w^2}
\left(\frac{\left(1+\lambda_{HV} \right)Q_d}{2} + \frac{1}{16} \right)
\right] 
\nonumber \\
C_+ & = & \frac{im_u V_{du}^\dagger}{4 \sqrt{2} m_W s_w} \left[
\frac{1}{c_w^2}
\left(
\frac{-1}{16} - \frac{\left(1+\lambda_{HV} \right)Q_u Q_d}{2}
\right)
- \frac{3}{32s_w^2}
\right. \nonumber\\
&-&\left.
 \frac{m_d^2}{16 s_w^2 m_W^2}
+\frac{I_{3d}}{c_w^2}
\left(\frac{\left(1+\lambda_{HV} \right)Q_u}{2} - \frac{1}{16} \right)
\right] 
\nonumber \\
\nonumber \\
%
%
\phi^+ d \bar u ~~:~~
C_- & = & - \frac{ im_u V_{ud}}{4 \sqrt{2} s_w m_W} \left[
\frac{1}{c_w^2}
\left(
\frac{1}{16} + \frac{\left(1+\lambda_{HV} \right)Q_u Q_d}{2}
\right)
+ \frac{3}{32 s_w^2}
\right. \nonumber\\
&+&\left.
 \frac{m_d^2}{16 s_w^2 m_W^2}
-\frac{I_{3d}}{c_w^2}
\left(\frac{\left(1+\lambda_{HV} \right)Q_u}{2} - \frac{1}{16} \right)
\right] \nonumber \\
C_+ & = & \frac{im_d V_{ud}}{4 \sqrt{2} m_W s_w} \left[
\frac{1}{c_w^2}
\left(
\frac{1}{16} + \frac{\left(1+\lambda_{HV} \right)Q_u Q_d}{2}
\right)
+ \frac{3}{32 s_w^2}
\right. \nonumber\\
&+&\left.
 \frac{m_u^2}{16 s_w^2 m_W^2}
-\frac{I_{3u}}{c_w^2}
\left(\frac{\left(1+\lambda_{HV} \right)Q_d}{2} + \frac{1}{16} \right)
\right] 
\nonumber \\
\nonumber \\
%
%
\phi^- \nu_{l} \bar{l} ~~:~~
C_- & = & -\frac{im_l}{4 \sqrt{2} m_W s_w} \left[
\frac{Q_l}{16 c_w^2}
- \frac{3}{32 s_w^2}
+\frac{I_{3\nu_l}}{c_w^2}
\left(\frac{\left(1+\lambda_{HV} \right)Q_l}{2} + \frac{1}{16} \right)
\right] \nonumber \\
C_+ & = & 0 
\nonumber \\
\nonumber \\
%
%
\phi^+ l \bar{\nu}_{l} ~~:~~
C_- & = & 0 \nonumber \\
C_+ & = & \frac{im_l}{4 \sqrt{2} m_W s_w} \left[
-\frac{Q_l}{16 c_w^2}
+ \frac{3}{32 s_w^2}
-\frac{I_{3\nu_l}}{ c_w^2}
\left(\frac{\left(1+\lambda_{HV} \right)Q_l}{2} + \frac{1}{16} \right)
\right]
\eqa

\subsubsection{{Vector-Fermion-Fermion effective vertices}}
The generic effective vertex is

\vspace{0.3cm}

%
\begin{center}
\fbox{
  \begin{picture}(230,65) 
    \SetOffset(-10,25)
    \SetWidth{0.5}
    \SetScale{0.5}
    \SetColor{Black}
    \Vertex(146,15){6.}
    \ArrowLine(200,57)(146,15)
    \ArrowLine(146,15)(200,-27)
    \Photon(72,15)(146,15){5.5}{4}
    \Text(30,7.5)[r]{\Black{$V_\mu$}}
    \Text(102, 30)[l]{$F_1$}
    \Text(102,-15)[l]{$\bar F_2$}
    \Text(115,7.5)[l]{$\displaystyle =\frac{i e^3}{\pi^2} (C_{-}\Omega^{-}+C_{+}\Omega^{+}) \gamma_\mu$}  
  \end{picture}
}
\end{center}
with the actual values of $V$, $F_1$, $\bar F_2$, $C_-$ and $C_+$
\bqa
%
%
A u \bar u ~~:~~
C_- &=& \frac{1}{4} \left[
\frac{\left(1+\lambda_{HV}\right)Q_u^3}{4 c_w^2} 
+\frac{m_u^2}{8 s_w^2 m_W^2} 
\left(\frac{1}{2}
\sum_{j=1}^3 \left(V_{u d_j} V_{d_j u}^\dagger  Q_{d_j}\right)
\right. \right. \nonumber \\
&+& \left. \left. \frac{Q_u}{4} + Q_u I_{3u}^2
\right)
\right] 
\nonumber \\
C_+ &=& \frac{1}{4} \left[
\frac{\left(1+\lambda_{HV}\right)Q_u^3}{4 c_w^2} - \frac{\left(1+\lambda_{HV}\right)Q_u^2 I_{3u}}{2c_w^2} +
\frac{\left(1+\lambda_{HV}\right)Q_u I_{3u}^2}{4 s_w^2 c_w^2}  \right.\nonumber \\
 & &+ \left.\frac{1}{4 s_w^2} 
 \left(\frac{1}{4 m_W^2}
\sum_{j=1}^3 \left(V_{u d_j} V_{d_j u}^\dagger m_{d_j}^2 Q_{d_j}\right)
 \right.\right.\nonumber \\
& & \left.\left. + \frac{m_u^2 Q_u \left(1+4 I_{3u}^2\right)}{8 m_W^2}
+ \sum_{j=1}^3 \left(V_{u d_j} V_{d_j u}^\dagger \left(
1 + Q_{d_j}
\right)\frac{\left(1+\lambda_{HV}\right)}{2}\right)
\right)
\right] 
\nonumber \\
\nonumber \\
%
%
A d \bar d ~~:~~
C_- &=& \frac{1}{4} \left[
\frac{\left(1+\lambda_{HV}\right)Q_d^3}{4 c_w^2} 
+\frac{m_d^2}{8 s_w^2 m_W^2} 
\left(\frac{1}{2}
\sum_{i=1}^3 \left(V_{u_i d} V_{d u_i}^\dagger  Q_{u_i}\right)
\right. \right. \nonumber \\
&+& \left. \left. \frac{Q_d}{4} + Q_d I_{3d}^2
\right)
\right] 
\nonumber \\
C_+ &=& \frac{1}{4} \left[
\frac{\left(1+\lambda_{HV}\right)Q_d^3}{4 c_w^2} - \frac{\left(1+\lambda_{HV}\right)Q_d^2 I_{3d}}{2c_w^2} +
\frac{\left(1+\lambda_{HV}\right)Q_d I_{3d}^2}{4 s_w^2 c_w^2} \right.\nonumber \\
 & &+\left.\frac{1}{4 s_w^2} 
\left(\frac{1}{4 m_W^2}
\sum_{i=1}^3 \left(V_{u_i d} V_{d u_i}^\dagger m_{u_i}^2 Q_{u_i}\right)
 \right.\right.\nonumber \\
& & \left.\left. 
+ \frac{m_d^2 Q_d \left(1+4 I_{3d}^2\right)}{8 m_W^2}
+ \sum_{i=1}^3 \left(V_{u_i d} V_{d u_i}^\dagger \left(
  Q_{u_i} -1
\right)\frac{\left(1+\lambda_{HV}\right)}{2}\right)
\right)
\right] 
\nonumber \\
\nonumber \\
%
%
A l \bar l ~~:~~
C_- &=& \frac{1}{4} \left[
\frac{\left(1+\lambda_{HV}\right)Q_l^3}{4 c_w^2} 
+\frac{m_l^2}{8 s_w^2 m_W^2} 
\left(\frac{Q_l}{4} + Q_l I_{3l}^2
\right)
\right] 
\nonumber \\
C_+ &=& \frac{1}{4} \left[
\frac{\left(1+\lambda_{HV}\right)Q_l^3}{4 c_w^2} 
- \frac{\left(1+\lambda_{HV}\right)Q_l^2 I_{3l}}{2c_w^2} 
+\frac{\left(1+\lambda_{HV}\right)Q_l I_{3l}^2}{4 s_w^2 c_w^2} \right.\nonumber \\
&& +\left.\frac{1}{4 s_w^2} 
\left(
\frac{m_l^2 Q_l \left(1+4 I_{3l}^2\right)}{8 m_W^2} 
-\frac{\left(1+\lambda_{HV}\right)}{2}\right)\right]  
\nonumber \\
\nonumber \\
%
%
A \nu_l \bar \nu_l ~~:~~
C_- &=& 0
\nonumber \\
C_+ &=& 
\frac{1}{32 s_w^2} 
\left[
\frac{m_{l}^2Q_{l}}{2 m_W^2}
+ \left(Q_{l}+1\right)\left(1+\lambda_{HV}\right)
\right] 
\nonumber \\
\nonumber \\
%
%
Z u \bar u ~~:~~
C_- &=& \frac{1}{8 c_w} \left\{
\frac{\left(1+\lambda_{HV}\right)Q_u^3 s_w}{2c_w^2} 
+\frac{m_u^2}{8 s_w  m_W^2} 
\left[
\sum_{j=1}^3 \left(V_{u d_j} V_{d_j u}^\dagger  
\left(Q_{d_j}-\frac{I_{3{d_j}}}{s_w^2}\right)
\right)
\right.\right.\nonumber \\
& & \left.\left.
+ \left(Q_u-\frac{I_{3u}}{s_w^2}\right)
\right]
\right\} 
\nonumber \\
C_+ &=& \frac{1}{8 c_w} \left\{
\frac{\left(1+\lambda_{HV}\right)Q_u^3 s_w}{2c_w^2} -  \frac{\left(1+\lambda_{HV}\right)Q_u^2 I_{3u} (1 + 2 s_w^2)}{ 2s_w c_w^2} \right.\nonumber \\
&&+3 \left.\frac{\left(1+\lambda_{HV}\right)Q_u I_{3u}^2}{2s_w c_w^2} -
\frac{\left(1+\lambda_{HV}\right)I_{3u}^3}{2s_w^3 c_w^2} \right. \nonumber \\
&+& 
\frac{1}{2s_w } 
\left[\frac{1}{4 m_W^2}
\left(
\sum_{j=1}^3 \left(V_{u d_j} V_{d_j u}^\dagger m_{d_j}^2 Q_{d_j}\right)
+ \frac{m_u^2 Q_u \left(1+4 I_{3u}^2\right)}{2}
\right)
 \right.\nonumber \\
&+& \left.\left.
 \sum_{j=1}^3 \left(V_{u d_j} V_{d_j u}^\dagger \frac{\left(1+\lambda_{HV}\right)}{2}
\left(Q_{d_j}- \frac{c_w^2 + I_{3d_j}}{s_w^2}
\right)
\right)
\right]
\right\} 
\nonumber \\
\nonumber \\
%
%
Z d \bar d ~~:~~
C_- &=& \frac{1}{8 c_w} \left\{
\frac{\left(1+\lambda_{HV}\right)Q_d^3 s_w}{2c_w^2} 
+\frac{m_d^2}{8 s_w  m_W^2} 
\left[
\sum_{i=1}^3 \left(V_{u_i d} V_{d u_i}^\dagger  
\left(Q_{u_i}-\frac{I_{3{u_i}}}{s_w^2}\right)
\right)
\right.\right.\nonumber \\
& & \left.\left.
+ \left(Q_d-\frac{I_{3d}}{s_w^2}\right)
\right]
\right\} 
\nonumber \\
C_+ &=& \frac{1}{16 c_w} \left\{\left(1+\lambda_{HV}\right)\left(
\frac{Q_d^3 s_w}{c_w^2} -  \frac{Q_d^2 I_{3d} (1 + 2 s_w^2)}{ s_w c_w^2} 
+3\frac{Q_d I_{3d}^2}{s_w c_w^2} -
\frac{I_{3d}^3}{s_w^3 c_w^2} \right)\right. \nonumber \\
&+& \left. \frac{1}{s_w} 
\left[\frac{1}{4 m_W^2}
\left(\sum_{i=1}^3 \left(V_{u_i d} V_{d u_i}^\dagger m_{u_i}^2 Q_{u_i}\right)
 \right.\right.\right.
\left. + \frac{m_d^2 Q_d \left(1+4 I_{3d}^2\right)}{2}
\right)
\nonumber \\
&+& \left.\left.
 \sum_{i=1}^3 \left(\frac{1+\lambda_{HV}}{2}\right)\left(V_{u_i d} V_{d u_i}^\dagger 
\left(Q_{u_i}+\frac{c_w^2 - I_{3u_i}}{s_w^2}\right)              
\right)
\right]
\right\} 
\nonumber \\
\nonumber \\
%
%
Z l \bar l ~~:~~
C_- &=& \frac{1}{8 c_w} \left\{
\frac{\left(1+\lambda_{HV}\right)Q_l^3 s_w}{2c_w^2} 
+\frac{m_l^2}{4 s_w  m_W^2} 
\left[
-\frac{1}{4s_w^2}
+ \frac{1}{2} \left(Q_l-\frac{I_{3l}}{s_w^2}\right)
\right]
\right\} \nonumber \\ 
C_+ &=& \frac{1}{16 c_w} \left\{\left(
\frac{Q_l^3 s_w}{c_w^2} -  \frac{Q_l^2 I_{3l} (1 + 2 s_w^2)}{ s_w c_w^2} \right.\right.\nonumber \\
&&+3 \left.\left.\frac{Q_l I_{3l}^2}{s_w c_w^2} -
\frac{I_{3l}^3}{s_w^3 c_w^2}\right)\left(1+\lambda_{HV}\right)
+\frac{1}{2 s_w } 
\left[
 \frac{m_l^2 Q_l}{2 m_W^2}
\right.\right. \nonumber \\
&&+ \left.\left.\frac{1}{s_w^2} \left(1+\lambda_{HV}\right)
\left(c_w^2 - I_{3\nu_{l}}\right)
\right]
\right\} 
\nonumber\\
\nonumber\\
%
%
Z \nu_l \bar \nu_l ~~:~~
C_- &=& 0
\nonumber \\
C_+ &=& \frac{1}{16 c_w} \left\{
-\frac{\left(1+\lambda_{HV}\right)I_{3\nu_l}^3}{s_w^3 c_w^2}
+\frac{1}{2 s_w} 
\left[\frac{m_{l}^2 Q_{l}}{2 m_W^2}
\right.\right.\nonumber \\
&& +\left. \left.\left(1+\lambda_{HV}\right)\left(Q_{l} - \frac{c_w^2 + I_{3l}}{s_w^2}\right)
\right]
\right\} 
\nonumber\\
\nonumber\\
%
%
W^- u \bar{d} ~~:~~
C_- & = &  0
\nonumber \\
C_+ &=& \frac{V_{du}^\dagger}{16\sqrt{2}s_w} 
\left[
\frac{Q_d I_{3u}+ Q_u I_{3d} - Q_u Q_d}{ c_w^2}
- \frac{1}{s_w^2}
+ \frac{1}{4 s_w^2 c_w^2}
\right] \left(1+\lambda_{HV}\right)
\nonumber\\
\nonumber\\
%
%
W^+ d \bar{u} ~~:~~
C_- & = & 0
\nonumber   \\
C_+ &=& \frac{V_{ud}}{16\sqrt{2}s_w} 
\left[
\frac{Q_d I_{3u}+ Q_u I_{3d} - Q_u Q_d}{ c_w^2}
- \frac{1}{s_w^2}
+ \frac{1}{4 s_w^2 c_w^2}
\right] \left(1+\lambda_{HV}\right)
\nonumber\\
\nonumber\\
%
%
\left.
\begin{tabular}{l}
$W^- \nu_l \bar{l}$ \\
$W^+ l \bar \nu_l$
\end{tabular}\right\}~~:~~
C_- & = & 0
\nonumber \\
C_+ &=& \frac{1}{16\sqrt{2} s_w} 
\left[
\frac{Q_l I_{3\nu_l}}{c_w^2}
- \frac{1}{s_w^2}
+ \frac{1}{4 s_w^2 c_w^2}
\right] \left(1+\lambda_{HV}\right)
\nonumber \\
\eqa
\subsubsection{{Scalar-Scalar-Scalar effective vertices}}
The generic effective vertex is

\vspace{0.3cm}

%
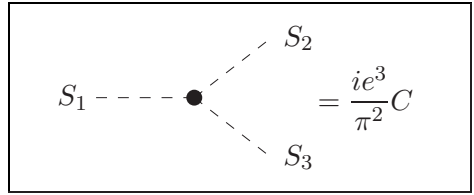
\begin{center}
\fbox{
  \begin{picture}(161,65) 
    \SetOffset(-10,25)
    \SetWidth{0.5}
    \SetScale{0.5}
    \SetColor{Black}
    \Vertex(146,15){6.0}
    \DashLine(72,15)(146,15){8}
    \DashLine(146,15)(200,57){8}
    \DashLine(200,-27)(146,15){8}
    \Text(33,7.5)[r]{\Black{$S_1$}}
    \Text(107, 30)[l]{$S_2$}
    \Text(107,-15)[l]{$S_3$}
    \Text(120,7.5)[l]{$\displaystyle =\frac{ie^3}{\pi^2} C $}
   \end{picture}
}
\end{center}
with the actual values of $S_1$, $S_2$, $S_3$, and $C$
\bqa
%
%
\left.
\begin{tabular}{l}
$HH\chi$        \\
$\chi\chi\chi$ \\
$\chi\phi^+\phi^-$
\end{tabular}\right\}~~:~~
C  & = &  0
\nonumber\\
\nonumber\\
HHH~~:~~C  & = & 
\frac{3}{32 s_w^3}\left[
\frac{1-4\lambda_{HV}}{2} m_W + \frac{1}{m_W^3} \left(
\sum_{i=1}^6 m_{l_i}^4 + N_{\mathrm{col}} \sum_{i=1}^6 m_{q_i}^4 
\right) \right. \nonumber \\
&+& \left.
\frac{1}{4}\left(1+\frac{1}{2 c_w^2}\right)\frac{m_H^2}{m_W}
+\frac{\left(1-4\lambda_{HV}\right) m_W}{4 c_w^4} 
\right] 
\nonumber\\
\nonumber\\
%
%
H\chi\chi~~:~~C  & = & 
\frac{1}{8 s_w^3}\left[
\frac{1-4\lambda_{HV}}{8} m_W + \frac{1}{4 m_W^3} \left(
\sum_{i=1}^6 m_{l_i}^4 
+ N_{\mathrm{col}} \sum_{i=1}^6 m_{q_i}^4  
\right) \right. \nonumber \\
&+& \left.
\frac{1}{16}\left(1+\frac{1}{2 c_w^2}\right)\frac{m_H^2}{m_W}
+\frac{\left(1-4\lambda_{HV}\right) m_W}{16 c_w^4} 
\right]
\nonumber\\
\nonumber\\
%
%
H\phi^+\phi^-~~:~~C  & = & 
\frac{1}{32 s_w^3}\left[
\frac{1}{m_W^3}\left(
\sum_{i=1}^3 m_{e_i}^4 
+ N_{\mathrm{col}} \sum_{i,j=1}^3 
\left(V_{u_id_j}V_{d_ju_i}^\dagger
(m_{u_i}^4+m_{d_j}^4)   
\right) \right)
\right.\nonumber\\
&+&\left. 
\frac{(1+2c_w^2)}{8 c_w^2}\frac{m_H^2}{m_W}
+\frac{3\left(1-4\lambda_{HV}\right)}{4}m_W 
+\frac{1-4\lambda_{HV}}{4} \frac{s_w^2 \left(1+c_w^2\right)}{c_w^4} m_W
\right]
\nonumber \\
\eqa
\subsubsection{{Vector-Scalar-Scalar effective vertices}}
The generic effective vertex is

\vspace{0.3cm}

%
\begin{center}
\fbox{
  \begin{picture}(210,70) 
    \SetOffset(-10,28)
    \SetWidth{0.5}
    \SetScale{0.5}
    \SetColor{Black}
    \Vertex(152,15){6.0}
    \Photon(72,15)(146,15){5.5}{4}
    \DashLine(146,15)(200,57){8}
    \DashLine(200,-27)(146,15){8}
    \LongArrow(180,62.3)(156,43) 
    \LongArrow(180,-31)(156,-12) 
    \Text(75,29)[lb]{{\Black{$p_1$}}}
    \Text(75,-13)[lt]{{\Black{$p_2$}}}
    \Text(33,7.5)[r]{\Black{$V_\mu$}}
    \Text(107, 30)[l]{$S_1$}
    \Text(107,-15)[l]{$S_2$}
    \Text(125,7.5)[l]{$\displaystyle = \frac{e^3}{\pi^2} C (p_1-p_2)_\mu$}
   \end{picture}
}
\end{center}
with the actual values of $V$, $S_1$, $S_2$, and $C$
\bqa
%
%
\left.
\begin{tabular}{l}
$AHH$       \\
$ZHH$       \\
$A\chi\chi$ \\
$Z\chi\chi$
\end{tabular}\right\}:~~C  & = & 0
\nonumber\\
\nonumber\\
%
%
A\chi H~~:~~C  & = &
\frac{5}{192 s_w^2}
\nonumber\\
\nonumber\\
%
%
Z\chi H~~:~~C  & = &
-\frac{1}{96 s_w c_w}\left[
\frac{1+ 2 c_w^2 + 20 c_w^4}{8 s_w^2 c_w^2}
+ \frac{1}{s_w^2 m_W^2}\left(
\sum_{i=1}^6 \left(m_{l_i}^2 +N_{\mathrm{col}} m_{q_i}^2 \right)
\right)
\right]
\nonumber\\
\nonumber\\
%
%
A\phi^+ \phi^-:~~C  & = &
\frac{i}{48 s_w^2}\left[
\frac{1+12c_w^2}{8c_w^2} 
+ \frac{1}{m_W^2} \left(-\sum_{i=1}^3 
\left(m_{e_i}^2 Q_{e_i}\right) \right. \right. \nonumber \\
&+& \left. \left. N_{\mathrm{col}} \sum_{i,j=1}^3 
\left(V_{u_id_j}V_{d_ju_i}^\dagger
(m_{u_i}^2 + m_{d_j}^2)
\right)
\right) 
\right]
\nonumber\\
\nonumber\\
%
%
Z\phi^+ \phi^-:~~C  & = &
\frac{i}{48 s_w c_w}\left\{
\frac{1-24c_w^4}{16 c_w^2 s_w^2} 
+ \frac{1}{m_W^2} \left(-\sum_{i=1}^3\left( 
m_{e_i}^2 \left(Q_{e_i} + \frac{I_{3\nu_{i}}}{s_w^2}\right)
\right) \right.\right.\nonumber\\
&+& \left.\left.
 N_{\mathrm{col}} \sum_{i,j=1}^3 
\left(V_{u_id_j}V_{d_ju_i}^\dagger \left[
(m_{u_i}^2 + m_{d_j}^2)
 + \frac{m_{u_i}^2 I_{3d_j} - m_{d_i}^2 I_{3u_i}}{s_w^2}
\right]
\right)
\right) 
\right\} 
\nonumber\\
\nonumber\\
%
%
\left.
\begin{tabular}{l}
$W^+\phi^- H$  \\ 
$W^- H \phi^+$
\end{tabular}\right\}:~~C  & = & 
\frac{i}{96 s_w^3}\left[
\frac{1+22c_w^2}{8 c_w^2} 
+ \frac{1}{m_W^2} \left(\sum_{i=1}^3 
m_{e_i}^2  
\right. \right. \nonumber \\
&+& \left. \left. N_{\mathrm{col}} \sum_{i,j=1}^3
\left(V_{u_id_j}V_{d_ju_i}^\dagger
(m_{u_i}^2 + m_{d_j}^2)
\right)
\right) 
\right]
\nonumber\\
\nonumber\\
%
%
\left.
\begin{tabular}{l}
$W^+\phi^- \chi$  \\ 
$W^-\phi^+ \chi$
\end{tabular}\right\}:~~C  & = & 
\frac{1}{48 s_w^3}\left[
-\frac{1+22c_w^2}{16 c_w^2} 
+ \frac{1}{m_W^2} \left(\sum_{i=1}^3 
\left(m_{e_i}^2  I_{3e_i}\right)
\right. \right. \nonumber \\
&-& \left. \left. N_{\mathrm{col}} \sum_{i,j=1}^3 
\left(V_{u_id_j}V_{d_ju_i}^\dagger
(m_{u_i}^2 I_{3u_i}- m_{d_j}^2 I_{3d_j})
\right)
\right) 
\right] 
\nonumber \\
\eqa
\subsubsection{{Scalar-Vector-Vector effective vertices}}
The generic effective vertex is

\vspace{0.3cm}

%
\begin{center}
\fbox{
  \begin{picture}(190,70) 
    \SetOffset(-10,28)
    \SetWidth{0.5}
    \SetScale{0.5}
    \SetColor{Black}
    \Vertex(146,15){6.0}
    \DashLine(72,15)(146,15){8}
    \Photon(146,15)(200,57){5.5}{4}
    \Photon(200,-27)(146,15){5.5}{4}
    \Text(33,7.5)[r]{\Black{$S$}}
    \Text(107, 30)[l]{$V_{1\mu}$}
    \Text(107,-15)[l]{$V_{2\nu}$}
    \Text(125,7.5)[l]{$\displaystyle = \frac{ie^3}{\pi^2} C g_{\mu\nu}$}
   \end{picture}
}
\end{center}
with the actual values of $S$, $V_1$, $V_2$ and $C$
\bqa
%
%
\left.
\begin{tabular}{l}
$\chi AA$       \\
$\chi AZ$       \\
$\chi ZZ$       \\
$\chi W^- W^+$
\end{tabular}\right\}:~~C  & = & 0
\nonumber\\
\nonumber\\
%
%
H AA ~~:~~ C &=&
- \frac{1}{8 s_w}
\left[\frac{1}{m_W}
\left(
\sum_{i=1}^6 \left(m_{l_i}^2 Q_{l_i}^2\right)
+ N_{\mathrm{col}} \sum_{i=1}^6 \left(m_{q_i}^2 Q_{q_i}^2\right)
\right) 
+ \frac{m_W}{2}
\right]
\nonumber\\
\nonumber\\
%
%
H AZ ~~:~~ C &=& 
\frac{1}{8 c_w}
\left\{\frac{1}{m_W}
\left[
\sum_{i=1}^6 \left(m_{l_i}^2 Q_{l_i} \left(\frac{I_{3l_i}}{2 s_w^2}-Q_{l_i}  \right)
\right) \right. \right. \nonumber \\
&+& \left. \left. N_{\mathrm{col}} \sum_{i=1}^6 \left(m_{q_i}^2 Q_{q_i} 
\left(\frac{I_{3q_i}}{2 s_w^2} - Q_{q_i}
\right)
\right)
\right] 
+ \frac{m_W \left(1+2 c_w^2\right)}{4 s_w^2}\right\} 
\nonumber \\  
\nonumber \\  
%
%
H ZZ ~~:~~ C &=& 
\frac{1}{8}
\left\{\frac{1}{m_W c_w^2}
\left[
\sum_{i=1}^6 \left(m_{l_i}^2 \left(
\frac{Q_{l_i} I_{3l_i}}{s_w}-Q_{l_i}^2 s_w  
-\frac{I_{3l_i}^2}{s_w^3}
\right)
\right) \right. \right. \nonumber \\
&+& \left. \left.  N_{\mathrm{col}} \sum_{i=1}^6 \left(m_{q_i}^2 \left(
\frac{Q_{q_i} I_{3q_i}}{s_w} - Q_{q_i}^2 s_w
-\frac{I_{3q_i}^2}{s_w^3}
\right)
\right)
\right] 
+ \frac{m_W \left(s_w^2 - 2\right)}{2 s_w^3}
\right\}
\nonumber \\  
\nonumber \\  
%
%
H W^-W^+ ~~:~~ C &=& 
-\frac{1}{8 s_w^3}
\left[\frac{1}{4 m_W}
\left(
\sum_{i=1}^3 m_{e_i}^2 
\right. \right. \nonumber\\
&+& \left. \left. N_{\mathrm{col}} \sum_{i,j=1}^3  
\left(V_{u_id_j}V_{d_ju_i}^\dagger 
\left(m_{u_i}^2 + m_{d_j}^2 \right) \right)
\right)
+ m_W
\right] 
\nonumber \\  
\nonumber \\  
%
%
\left.
\begin{tabular}{l}
$\phi^- A W^+$       \\
$\phi^+ W^- A$       
\end{tabular}\right\}:~~C  & = & 
\frac{1}{32 s_w^2} K
\nonumber\\
\nonumber\\
%
%
\left.
\begin{tabular}{l}
$\phi^- Z W^+$       \\
$\phi^+ W^- Z$       
\end{tabular}\right\}:~~C  & = & 
\frac{1}{32 s_w c_w} K
\eqa
where
\bqa
K &=& m_W +
\frac{N_{\mathrm{col}}}{m_W}
 \sum_{i,j=1}^3 \left(V_{u_id_j}V_{d_ju_i}^\dagger 
\left(Q_{u_i} m_{d_j}^2 - Q_{d_j} m_{u_i}^2 \right)\right)
\eqa

\subsubsection{{Vector-Vector-Vector effective vertices}}
The generic effective vertex is

\vspace{0.3cm}

%
\begin{center}
\fbox{
  \begin{picture}(390,80) 
    \SetOffset(-6,34)
    \SetWidth{0.5}
    \SetScale{0.5}
    \SetColor{Black}
    \Vertex(146,15){6.0}
    \Photon(72,15)(146,15){5.5}{4}
    \Photon(146,15)(200,57){5.5}{4}
    \Photon(200,-27)(146,15){5.5}{4}
    \LongArrow(85,28)(120,28)
    \Text(52,18)[b]{{\Black{$p_1$}}}
    \LongArrow(180,62.3)(156,43) 
    \LongArrow(180,-31)(156,-12) 
    \Text(75,29)[lb]{{\Black{$p_2$}}}
    \Text(75,-13)[lt]{{\Black{$p_3$}}}
    \Text(33,7.5)[r]{\Black{$V_{1\alpha}$}}
    \Text(107, 30)[l]{$V_{2\mu}$}
    \Text(107,-15)[l]{$V_{3\nu}$}
    \Text(125,7.5)[l]{$\displaystyle = \frac{ie^3}{\pi^2}\,C\, 
\left[g_{\alpha\mu} (p_2-p_1)_\nu
     + g_{\mu\nu} (p_3-p_2)_\alpha  
     + g_{\nu\alpha}(p_1-p_3)_\mu 
  \right]$}
\end{picture}
 }
\end{center}
with the actual values of $V_1$, $V_2$, $V_3$ and $C$
\bqa
%
%
\left.
\begin{tabular}{l}
$AAA$       \\
$AAZ$       \\
$AZZ$       \\
$ZZZ$       \\
\end{tabular}\right\}:~~C  & = & 0
\nonumber\\
\nonumber\\
%
%
AW^+W^- ~~:~~ C &=& K
\nonumber \\
\nonumber \\
%
%
ZW^+W^- ~~:~~ C &=& -\frac{c_w}{s_w} K
\eqa
where
\bqa
K &=&   \frac{7+4\lambda_{HV}}{96 s_w^2} 
+\frac{1}{48 s_w^2}
\left[ \sum_{i=1}^3 1
+ N_{\mathrm{col}}\sum_{i,j=1}^{3} 
\left(V_{u_id_j}V_{d_ju_i}^\dagger \right)
\right]
\eqa
\subsection{\underline{Electroweak effective vertices with 4 external legs}}
In this section, we give all possible contributing 
4-point ${\rm R_2}$ effective vertices.
\subsubsection{{Scalar-Scalar-Scalar-Scalar effective vertices}}
The generic effective vertex is

\vspace{0.3cm}

\begin{center}
\fbox{
  \begin{picture}(130,73)
    \SetOffset(-8,31.5)
    \SetWidth{0.5}
    \SetScale{0.5}
    \SetColor{Black}
    \DashLine(57,-36)(161,54){8}
    \DashLine(57,54)(161,-36){8}
    \Vertex(110,9){6.0}
    \Text(26,-18)[tr]{$S_1$}
    \Text(26, 27)[br]{$S_2$}
    \Text(84, 27)[bl]{$S_3$}
    \Text(84,-18)[tl]{$S_4$}
    \Text(95,5)[l]{$\displaystyle =  \frac{ie^4}{\pi^2} \,C $}
  \end{picture}
}
\end{center}
with the actual values of $S_1$, $S_2$, $S_3$, $S_4$ and $C$
\bqa
%
%
\left.
\begin{tabular}{l}
$HHH\chi      $       \\
$H\chi\chi\chi$       \\
$H\chi\phi^-\phi^+$   \\
\end{tabular}\right\}:~~C  & = & 0
\nonumber\\
\nonumber\\
%
%
\left.
\begin{tabular}{l}
$HHHH$              \\
$\chi\chi\chi\chi$  \\
\end{tabular}\right\}:~~C  & = & \frac{1}{64 s_w^4} K_1
\nonumber\\
\nonumber\\
%
%
HH\chi\chi ~~:~~ C &=& 
\frac{1}{192 s_w^4} K_1
\nonumber\\
\nonumber\\
%
%
\left.
\begin{tabular}{l}
$HH\phi^-\phi^+$         \\
$\chi\chi\phi^-\phi^+$  \\
\end{tabular}\right\}:~~C  & = & \frac{1}{64 s_w^4} K_2
\nonumber\\
\nonumber\\
%
%
\phi^-\phi^+\phi^-\phi^+  ~~:~~ C &=& 
\frac{1}{32 s_w^4} K_3
\eqa
where
\bqa
K_1 &=&
\frac{1}{m_W^2}\left[
\frac{5}{m_W^2}
\sum_{i=1}^{6}  \left (m_{l_i}^4
+N_{\mathrm{col}} m_{q_i}^4 \right )  
+\frac{3}{2}m_H^2\left(1+\frac{1}{2 c_w^2}\right)
\right]
+\frac{1-12\lambda_{HV}}{2}
\left(1+\frac{1}{2 c_w^4}\right) 
\nonumber \\
\nonumber \\
K_2 &=&
\frac{1}{m_W^2}\left[
\frac{5}{3 m_W^2}\left(
\sum_{i=1}^{3} m_{e_i}^4
+N_{\mathrm{col}}\sum_{i,j=1}^{3} V_{u_id_j}V_{d_ju_i}^\dagger  
\left(m_{u_i}^4 + m_{d_j}^4\right)\right) 
+\frac{1}{2}m_H^2\left(1+\frac{1}{2 c_w^2}\right)
\right]
\nonumber\\ 
&+&\frac{1-12\lambda_{HV}}{4}\left(1+\frac{s_w^2}{3c_w^2}\left(1+\frac{1}{c_w^2}\right)\right)
\nonumber \\
\nonumber \\
K_3 &=&
\frac{1}{m_W^2}\left[\frac{5}{3 m_W^2}\left(
\sum_{i=1}^{3}  m_{e_i}^4
+N_{\mathrm{col}}\sum_{i,j,k,l=1}^{3}
\left(V_{u_id_j}V_{d_ju_k}^\dagger V_{u_kd_l}V_{d_lu_i}^\dagger  
\left(m_{u_i}^2m_{u_k}^2 + m_{d_j}^2m_{d_l}^2\right)\right)\right) 
\right.\nonumber\\
&+& \left.
\frac{1}{2}m_h^2\left(1+\frac{1}{2 c_w^2}\right)
\right]
+\left( \left(\frac{1}{4}-3\lambda_{HV}\right)
\left(1+s_w^4\right) 
\right. \nonumber \\
&+&  \left.\left(\frac{1}{6}-2\lambda_{HV}\right) \left(s_w^2+\frac{2s_w^6}{c_w^2}\right)+\left(\frac{1}{12}-\lambda_{HV}\right)\frac{s_w^8}{c_w^4}\right)
\nonumber \\
\eqa
\subsubsection{{Vector-Vector-Vector-Vector effective vertices}}
The generic effective vertex is

\vspace{0.3cm}

\begin{center}
\fbox{
  \begin{picture}(300,73)
    \SetOffset(0,31.5)
    \SetWidth{0.5}
    \SetScale{0.5}
    \SetColor{Black}
    \Photon(57,-36)(161,54){5.5}{8}
    \Photon(57,54)(161,-36){5.5}{8}
    \Vertex(110,9){6.0}
    \Text(26,-18)[tr]{$V_{1\alpha}$}
    \Text(26, 27)[br]{$V_{2\beta}$}
    \Text(84, 27)[bl]{$V_{3\mu}$}
    \Text(84,-18)[tl]{$V_{4\nu}$}
    \Text(95,5)[l]{$\displaystyle =
\frac{ie^4}{\pi^2} [C_1 g_{\alpha\beta}g_{\mu\nu}+
C_2 g_{\alpha\mu}g_{\beta\nu}+
C_3 g_{\alpha\nu}g_{\beta\mu}]
 $}
  \end{picture}
}
\end{center}
with the actual values of $V_1$, $V_2$, $V_3$, $V_4$ 
$C_1$,  $C_2$ and $C_3$ 
\bqa
%
%
AAAA ~~:~~ C_1 &=& 
\frac{1}{12}\left(-1+\sum_{i=1}^{6} Q_{l_i}^4
+N_{\mathrm{col}}\sum_{i=1}^{6} Q_{q_i}^4\right)
\nonumber \\
 C_2 &=& C_1
\nonumber \\
 C_3 &=& C_1
\nonumber\\
\nonumber\\
%
%
AAAZ ~~:~~ C_1 &=& 
\frac{1}{12}
\left[
\frac{c_w}{s_w}+\sum_{i=1}^{6}\left(\frac{s_w}{c_w} Q_{l_i}^4 -
\frac{1}{2 s_w c_w} Q_{l_i}^3 I_{3l_i}\right) \right. \nonumber \\
&+& \left. N_{\mathrm{col}}\sum_{i=1}^{6}\left(\frac{s_w}{c_w} Q_{q_i}^4 -
\frac{1}{2 s_w c_w} Q_{q_i}^3 I_{3q_i}
\right)
\right]
\nonumber \\
 C_2 &=& C_1
\nonumber \\
 C_3 &=& C_1
\nonumber\\
\nonumber\\
%
%
AAZZ ~~:~~ C_1 &=& 
\frac{1}{12}
\left[- \frac{c_w^2}{s_w^2} 
+ \frac{1}{2}\sum_{i=1}^{6}\left(
\frac{s_w^2}{c_w^2} Q_{l_i}^4 + \left(
\frac{s_w}{c_w}Q_{l_i}^2 - \frac{1}{s_w c_w}
Q_{l_i} I_{3l_i}\right)^2
\right) \right.\nonumber\\
&+ &\left. 
  \frac{N_{\mathrm{col}}}{2}\sum_{i=1}^{6}\left(
\frac{s_w^2}{c_w^2} Q_{q_i}^4 + \left(
\frac{s_w}{c_w}Q_{q_i}^2 - \frac{1}{s_w c_w}
Q_{q_i} I_{3q_i}\right)^2
\right)
\right] 
\nonumber \\
 C_2 &=& C_1
\nonumber \\
 C_3 &=& C_1
\nonumber\\
\nonumber\\
%
%
AZZZ ~~:~~ C_1 &=& 
\frac{1}{12}
\left[
\frac{c_w^3}{s_w^3}+
\sum_{i=1}^{6}\left(
\frac{s_w^3}{c_w^3} Q_{l_i}^4
-\frac{3}{2}\frac{s_w}{c_w^3}Q_{l_i}^3 I_{3l_i} 
\right. \right. \nonumber \\
&+& \left. \left. \frac{3}{2}\frac{1}{s_w c_w^3}Q_{l_i}^2 I_{3l_i}^2 
- \frac{1}{2 s_w^3 c_w^3} Q_{l_i} I_{3l_i}^3
\right)
\right. \nonumber \\
&+&  N_{\mathrm{col}}\sum_{i=1}^{6}\left(
\frac{s_w^3}{c_w^3} Q_{q_i}^4
-\frac{3}{2}\frac{s_w}{c_w^3}Q_{q_i}^3 I_{3q_i} \right. \nonumber \\
&+& \left. \left. \frac{3}{2}\frac{1}{s_w c_w^3}Q_{q_i}^2 I_{3q_i}^2 
- \frac{1}{2 s_w^3 c_w^3} Q_{q_i} I_{3q_i}^3
\right)
\right]
\nonumber \\
 C_2 &=& C_1
\nonumber \\
 C_3 &=& C_1
\nonumber\\
\nonumber\\
%
%
ZZZZ ~~:~~ C_1 &=&
\frac{1}{12}
\left[ -\frac{c_w^4}{s_w^4} 
+ \sum_{i=1}^{6} \left(
\frac{s_w^4}{c_w^4} Q_{l_i}^4 
- 2 \frac{s_w^2}{c_w^4} Q_{l_i}^3 I_{3l_i} 
\right. \right. \nonumber \\
&+& \left. \left. 
 \frac{3}{c_w^4} Q_{l_i}^2 I_{3l_i}^2
- \frac{2}{s_w^2 c_w^4} Q_{l_i} I_{3l_i}^3
+ \frac{1}{2 s_w^4 c_w^4} I_{3l_i}^4 
\right)
\right.\nonumber\\
&+&\left.
  N_{\mathrm{col}}\sum_{i=1}^{6} \left(
\frac{s_w^4}{c_w^4} Q_{q_i}^4 
- 2 \frac{s_w^2}{c_w^4} Q_{q_i}^3 I_{3q_i} 
\right. \right. \nonumber \\
&+& \left. \left. 
 \frac{3}{c_w^4} Q_{q_i}^2 I_{3q_i}^2
- \frac{2}{s_w^2 c_w^4} Q_{q_i} I_{3q_i}^3
+ \frac{1}{2 s_w^4 c_w^4} I_{3q_i}^4 
\right)
\right]
\nonumber \\
 C_2 &=& C_1
\nonumber \\
 C_3 &=& C_1
\nonumber\\
\nonumber\\
%
%
AAW^-W^+ ~~:~~ C_1 &=&
 \frac{1}{16 s_w^2} \left[ \frac{10+4\lambda_{HV}}{3} + \sum_{i=1}^3
1
+ \frac{25}{27} N_{\mathrm{col}}\sum_{i,j=1}^3\left(V_{u_id_j}V_{d_ju_i}^\dagger 
\right)
\right]
\nonumber \\
C_2 &=&
- \frac{1}{16 s_w^2} \left[ \frac{7+2\lambda_{HV}}{3} + \frac{1}{3}\sum_{i=1}^3
1 
+\frac{11}{27} N_{\mathrm{col}}\sum_{i,j=1}^3 \left(V_{u_id_j}V_{d_ju_i}^\dagger 
\right)
\right]
\nonumber \\
 C_3 &=& C_2
\nonumber\\
\nonumber\\
%
%
AZW^-W^+ ~~:~~ C_1 &=&
\frac{1}{16 s_w c_w} \left[
- \frac{\left(10+4\lambda_{HV}\right) c_w^2}{3 s_w^2} +    
\left( 1 - \frac{11}{12 s_w^2} \right) \sum_{i=1}^3 1
\right. \nonumber \\
&+&\left. 
N_{\mathrm{col}}\sum_{i,j=1}^3 \left(V_{u_id_j}V_{d_ju_i}^\dagger 
\left( 
\frac{25}{27}-\frac{11}{12s_w^2}
\right)
\right)
\right] 
\nonumber \\
C_2 &=&
\frac{1}{16 s_w c_w} \left[
\frac{7+2\lambda_{HV}}{3} \frac{c_w^2}{s_w^2}    
+\left(
\frac{5}{12 s_w^2}- \frac{1}{3}
\right) \sum_{i=1}^3 1
%
%
\right. \nonumber \\
&+&\left. 
N_{\mathrm{col}}\sum_{i,j=1}^3 \left(V_{u_id_j}V_{d_j u_i}^\dagger 
\left( 
\frac{5}{12 s_w^2}-\frac{11}{27}
\right)
\right)
\right] 
\nonumber \\
 C_3 &=& C_2
\nonumber\\
\nonumber\\
%
%
ZZW^-W^+ ~~:~~ C_1 &=&
\frac{\left(5+2\lambda_{HV}\right) c_w^2}{24 s_w^4}+ \frac{1}{16 c_w^2}  
\left[
 \left(
1-\frac{11}{6s_w^2}+\frac{11}{12s_w^4}
\right)  \sum_{i=1}^3 1
 \right.\nonumber\\
&+& \left.   N_{\mathrm{col}}
\left(
\frac{25}{27}-\frac{11}{6s_w^2}+\frac{11}{12s_w^4}
\right)
\sum_{i,j=1}^3 \left(
V_{u_i d_j}V_{d_j u_i}^\dagger
\right) \right]
\nonumber \\
C_2 &=&
-\frac{\left(7+2\lambda_{HV}\right) c_w^2}{48 s_w^4}
+ \frac{1}{16 c_w^2}  
\left[
\left(
-\frac{1}{3}+\frac{5}{6s_w^2}-\frac{5}{12s_w^4}
\right) \sum_{i=1}^3 1
 \right.\nonumber\\
&+& \left.   N_{\mathrm{col}}
\left(
-\frac{11}{27}+\frac{5}{6s_w^2}-\frac{5}{12s_w^4}
\right)
\sum_{i,j=1}^3 \left(
V_{u_i d_j}V_{d_j u_i}^\dagger
\right) \right]
\nonumber \\
 C_3 &=& C_2
\nonumber\\
\nonumber\\
%
%
W^-W^+W^-W^+ ~~:~~ C_1 &=&
\frac{1}{16 s_w^4} \left[
\frac{3+2\lambda_{HV}}{3}+\frac{1}{2}\sum_{i=1}^3 1 
\right. \nonumber \\
&+& \left . 
\frac{N_{\mathrm{col}}}{2}\sum_{i,j,k,m=1}^3 
\left(
V_{u_id_j}V_{d_ju_k}^\dagger V_{u_kd_m}V_{d_mu_i}^\dagger 
\right)
\right] 
\nonumber \\
C_2 &=&
-\frac{1}{8 s_w^4} \left[
\frac{7+2\lambda_{HV}}{3}+\frac{5}{12}\sum_{i=1}^3 1 
\right. \nonumber \\
&+& \left. \frac{5}{12} N_{\mathrm{col}}
\sum_{i,j,k,m=1}^3  
\left(
V_{u_id_j}V_{d_ju_k}^\dagger V_{u_kd_m}V_{d_mu_i}^\dagger 
\right)
\right]
\nonumber \\
 C_3 &=& C_1
\eqa
\subsubsection{{Scalar-Scalar-Vector-Vector effective vertices}}
The generic effective vertex is

\vspace{0.3cm}

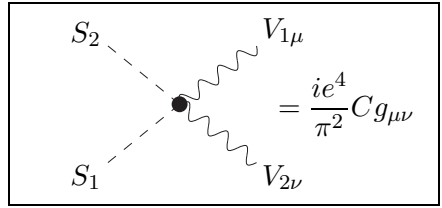
\begin{center}
\fbox{ 
 \begin{picture}(149,70)
    \SetOffset(-10,25)
    \SetWidth{0.5}
    \SetScale{0.5}
    \SetColor{Black}
    \DashLine(135,20)(81,-24){8}
    \DashLine(135,20)(81,64){8}
    \Photon(135,20)(193,64){5.5}{4}
    \Photon(135,20)(193,-24){5.5}{4}
    \Text(38,-13)[tr]{$S_{1}$}
    \Text(38, 32)[br]{$S_{2}$}
    \Text(99, 32)[bl]{$V_{1\mu}$}
    \Text(99,-13)[tl]{$V_{2\nu}$}
    \Vertex(135,20){6}
    \Text(105,10)[l]{$\displaystyle = \frac{ie^4}{\pi^2} C g_{\mu\nu}$}
  \end{picture}
}
\end{center}
with the actual values of $S_1$, $S_2$, $V_1$, $V_2$  and $C$ 
\bqa
%
%
\left.
\begin{tabular}{l}
$H\chi AA $       \\
$H\chi AZ $       \\
$H\chi ZZ $       \\
$H\chi W^+W^- $      
\end{tabular}\right\}:~~C  & = & 0
\nonumber\\
\nonumber\\
%
%
\left.
\begin{tabular}{l}
$H H       AA$       \\
$\chi \chi AA$       
\end{tabular}\right\}:~~C  & = & 
\frac{1}{16 s_w^2}\left\{\frac{1}{12}
-\frac{1}{m_W^2}\left[
\sum_{i=1}^{6} \left(Q_{l_i}^2 m_{l_i}^2\right)
+N_{\mathrm{col}}\sum_{i=1}^{6} \left(Q_{q_i}^2 m_{q_i}^2\right) 
\right]
\right\}
\nonumber\\
\nonumber\\
%
%
\left.
\begin{tabular}{l}
$H H       AZ$       \\
$\chi \chi AZ$       
\end{tabular}\right\}:~~C  & = & 
\frac{1}{16 s_w}\left\{\frac{4 + s_w^2}{12 s_w^2 c_w}
+\frac{1}{m_W^2 c_w}\left[
\sum_{i=1}^{6} \left(Q_{l_i} m_{l_i}^2
\left(\frac{I_{3l_i}}{2 s_w^2}-Q_{l_i}\right)
\right) \right.\right. \nonumber\\
&+& \left.\left.
N_{\mathrm{col}}\sum_{i=1}^{6} \left(Q_{q_i} m_{q_i}^2
\left(\frac{I_{3q_i}}{2 s_w^2}-Q_{q_i}\right)
\right) 
\right]
\right\}
\nonumber\\
\nonumber\\
%
%
\left.
\begin{tabular}{l}
$H H       ZZ$       \\
$\chi \chi ZZ$       
\end{tabular}\right\}:~~C  & = & 
 -\frac{1}{16 c_w^2}\left\{\frac{1 + 2 c_w^2 + 40  c_w^4 - 4  c_w^6}
{48 s_w^4 c_w^2} \right. \nonumber \\
&+& \left. \frac{1}{m_W^2}\left[
\sum_{i=1}^{6} \left(m_{l_i}^2
\left(Q_{l_i}^2 +\frac{4 I_{3l_i}^2}{3 s_w^4}-
\frac{Q_{l_i} I_{3l_i}}{s_w^2}\right)
\right) \right.\right.\nonumber\\
&+& \left.\left.
N_{\mathrm{col}}\sum_{i=1}^{6} \left(m_{q_i}^2
\left(Q_{q_i}^2 + \frac{ 4 I_{3q_i}^2}{3 s_w^4}
-\frac{Q_{q_i} I_{3q_i}}{s_w^2}\right)
\right) 
\right]
\right\}
\nonumber\\
\nonumber\\
%
%
\left.
\begin{tabular}{l}
$H H       W^-W^+$       \\
$\chi \chi W^-W^+$       
\end{tabular}\right\}:~~C  & = & 
 -\frac{1}{48 s_w^4}\left\{\frac{1 + 38 c_w^2}{16 c_w^2}
\right. \nonumber \\
&+& \left. \frac{1}{m_W^2}\left[
\sum_{i=1}^{3} m_{e_i}^2 
+ N_{\mathrm{col}}\sum_{i,j=1}^{3} 
\left(V_{u_id_j}V_{d_ju_i}^\dagger
\left(m_{u_i}^2 + m_{d_j}^2\right)
\right) 
\right]
\right\}
\nonumber\\
\nonumber\\
%
%
\left.
\begin{tabular}{l}
$ H  \phi^+ W^- A  $       \\
$\phi^-  H A W^+   $       
\end{tabular}\right\}:~~C  & = & K_1 
\nonumber\\
\nonumber\\
%
%
 \chi\phi^+ W^- A  ~~:~~ C &=&
-i K_1
\nonumber\\
\nonumber\\       
%
%
\phi^-  \chi A W^+  ~~:~~ C &=&
i K_1
\nonumber\\
\nonumber\\
%
%
\left.
\begin{tabular}{l}
$ H  \phi^+ W^- Z  $       \\
$\phi^-  H Z W^+   $       
\end{tabular}\right\}:~~C  & = & K_2 
\nonumber\\
\nonumber\\
%
%
 \chi\phi^+ W^- Z  ~~:~~ C &=&
- i K_2      
\nonumber\\
\nonumber\\
%
%
\phi^-  \chi Z W^+  ~~:~~ C &=&
i K_2
\nonumber\\
\nonumber\\
%
%
\phi^- \phi^+ A A  ~~:~~ C &=&
 -\frac{1}{12 s_w^2}\left\{\frac{1 + 21 c_w^2}{16 c_w^2}
+\frac{1}{m_W^2}
\left[
\sum_{i=1}^{3} 
m_{e_i}^2
\right.\right. \nonumber\\
&+&\left.\left.  
\frac{5}{6} N_{\mathrm{col}}\sum_{i,j=1}^{3} 
\left(V_{u_id_j}V_{d_ju_i}^\dagger
(m_{u_i}^2 + m_{d_j}^2)
\right) 
\right]
\right\}
\nonumber\\
\nonumber\\
%
%
 \phi^- \phi^+ A Z ~~:~~ C &=&
\frac{1}{12 s_w c_w}\left\{\frac{42 c_w^4-10 c_w^2-1}{32 s_w^2 c_w^2}
\right. \nonumber  \\
&-& \left. \frac{1}{m_W^2}
\left[
\sum_{i=1}^{3} 
\left(m_{e_i}^2 Q_{e_i}
\left(Q_{e_i}
+\frac{5}{8}\frac{I_{3\nu_i}}{s_w^2} 
\right)
\right)
\right.\right. \nonumber\\
&+&\left.\left.  
 N_{\mathrm{col}}\sum_{i,j=1}^{3} 
\left[V_{u_id_j}V_{d_ju_i}^\dagger
\left(
m_{u_i}^2
\left(\frac{5}{6} - 
\frac{I_{3d_i}}{s_w^2}
\left(Q_{d_j} -\frac{5}{8}Q_{u_i}\right)
\right)
\right.\right.\right.\right.\nonumber\\
&+& \left.\left.\left.\left.
 m_{d_j}^2 
\left(\frac{5}{6} - 
\frac{I_{3u_i}}{s_w^2}
\left(Q_{u_i} -\frac{5}{8}Q_{d_j}\right)\right)
\right)
\right] 
\right]
\right\}
\nonumber\\
\nonumber\\
%
%
 \phi^- \phi^+ Z Z ~~:~~ C &=&
\frac{1}{12 c_w^2}\left\{\frac{ 
- 1 + 2 c_w^2 + 44 c_w^4 - 84 c_w^6}{64 s_w^4 c_w^2}
\right. \nonumber \\
&-& \left. \frac{1}{m_W^2}
\left[
\sum_{i=1}^{3} 
\left(m_{e_i}^2 
\left(Q_{e_i}^2  
+\frac{5}{4}\frac{Q_{e_i}I_{3\nu_i}}{s_w^2} 
+\frac{I_{3\nu_i}^2}{s_w^4}
\right)
\right)
\right.\right. \nonumber\\
&+&\left.\left.  
 N_{\mathrm{col}}\sum_{i,j=1}^{3} 
\left[V_{u_id_j}V_{d_ju_i}^\dagger
\left(
m_{u_i}^2
\left(\frac{5}{6} - 
\frac{I_{3d_i}}{s_w^2}
\left(2Q_{d_j} -\frac{5}{4}Q_{u_i}\right)
+\frac{I_{3d_i}^2}{s_w^4}
\right)
\right.\right.\right.\right.\nonumber\\
&+& \left.\left.\left.\left.
 m_{d_j}^2 
\left(\frac{5}{6}- 
\frac{I_{3u_i}}{s_w^2}
\left(2 Q_{u_i} -\frac{5}{4}Q_{d_j}\right)\right)
+\frac{I_{3u_i}^2}{s_w^4}
\right)
\right] 
\right]
\right\}
\nonumber\\
\nonumber\\
%
%
\phi^-\phi^+ W^- W^+ ~~:~~ C &=&
- \frac{1}{48 s_w^4}\left\{
\frac{1}{m_W^2}\left[\left(
\sum_{i=1}^{3}  m_{e_i}^2 \right. \right. \right. \nonumber \\
&+& \left. \left. \left. N_{\mathrm{col}}\sum_{i,j,k,l=1}^{3}
\left(V_{u_id_j}V_{d_ju_k}^\dagger V_{u_kd_l}V_{d_lu_i}^\dagger
\left(m_{u_i} m_{u_k} + m_{d_j} m_{d_l}\right)\right)\right)\right]
\right.\nonumber\\
&+& \left.
\frac{38c_w^2+1}
{16c_w^2}
\right\}
\nonumber \\
\eqa
with
\bqa
K_1 &=& \frac{1}{24 s_w^3}\left\{\frac{1 + 22 c_w^2}{32 c_w^2}
+K \right\}\nonumber \\ 
K_2 &=& \frac{1}{24 s_w^2 c_w}\left\{\frac{1 + 21 c_w^2 -22 c_w^4}{32 c_w^2s_w^2}
+K \right\} \nonumber \\ 
K &=& \frac{1}{8 m_W^2} \left[
\sum_{i=1}^{3} m_{e_i}^2  
+N_{\mathrm{col}} \sum_{i,j=1}^{3} 
\left(V_{u_id_j}V_{d_ju_i}^\dagger
\left( 3 m_{d_j}^2 + 2 m_{u_i}^2
\right)
\right) 
\right]
\eqa
\subsection{\underline{Mixed Electroweak/QCD corrections}}
In~\cite{qcdrational}, all mixed ${\rm R_2}$ QCD/Electroweak vertices 
with internal QCD particle and external weak fields are presented. 
For completeness, we give here the only contributing Mixed Electroweak/QCD ${\rm R_2}$ 
effective vertex, with internal EW particles and external colored states.
\subsubsection{{Gluon-Quark-Quark effective vertex}}
The generic effective vertex is

\vspace{0.3cm}

%
\begin{center}
\fbox{
  \begin{picture}(250,65) 
    \SetOffset(-10,25)
    \SetWidth{0.5}
    \SetScale{0.5}
    \SetColor{Black}
    \Vertex(146,15){6.}
    \ArrowLine(200,57)(146,15)
    \ArrowLine(146,15)(200,-27)
    \Gluon(72,15)(146,15){5.5}{5}
    \Text(30,7.5)[r]{\Black{$G^a_\mu$}}
    \Text(102, 30)[l]{$Q_l$}
    \Text(102,-15)[l]{$\bar Q_k$}
    \Text(115,7.5)[l]{$\displaystyle=
\frac{i g_s e^2}{\pi^2} t_{k l}^a
(C_{-}\Omega^{-}+C_{+}\Omega^{+}) \gamma_\mu $}  
  \end{picture}
}
\end{center}
with the actual values of $Q$, $\bar Q$, $C_-$ and $C_+$
\bqa
%
%
u \bar u ~~:~~ C_- &=&
\frac{1}{16} \left[
(1+\lambda_{HV}) \frac{Q_u^2}{c_w^2}
+\frac{m_u^2}{2 s_w^2 m_W^2}
\left(\frac{1}{2}
\sum_{j=1}^3 \left(V_{u d_j} V_{d_j u}^\dagger \right)
+ \frac{1}{4} + I_{3u}^2
\right)
\right]
\nonumber\\
C_+ &=& 
\frac{1}{16} \left[
\left(1+\lambda_{HV}\right)
\left(
\frac{1}{c_w^2}\left(Q_u^2+\frac{I_{3u}^2}{s_w^2} -2 Q_u I_{3u}
\right)
+\frac{1}{2 s_w^2} 
\sum_{j=1}^3 \left(V_{u d_j} V_{d_j u}^\dagger\right)
\right)
\right.\nonumber \\
& & \left.
+ \frac{1}{2 m_W^2 s_w^2}
\left(
\frac{1}{2}
\sum_{j=1}^3 \left(V_{u d_j} V_{d_j u}^\dagger m_{d_j}^2 \right)
+ m_u^2 \left(\frac{1}{4} + I_{3u}^2\right)
\right)
\right] 
\nonumber\\
\nonumber\\
%
%
d \bar d ~~:~~ C_- &=&
\frac{1}{16} \left[
(1+\lambda_{HV}) \frac{Q_d^2}{c_w^2}
+\frac{m_d^2}{2 s_w^2 m_W^2}
\left(\frac{1}{2}
\sum_{i=1}^3 \left(V_{u_i d} V_{d u_i}^\dagger \right)
+ \frac{1}{4} + I_{3d}^2
\right)
\right]
\nonumber\\
C_+ &=& 
\frac{1}{16} \left[
\left(1+\lambda_{HV}\right)
\left(
\frac{1}{c_w^2}\left(Q_d^2+\frac{I_{3d}^2}{s_w^2} -2 Q_d I_{3d}
\right)
+\frac{1}{2 s_w^2}
\sum_{i=1}^3 \left(V_{u_i d} V_{d u_i}^\dagger\right)
\right)
\right.\nonumber \\
& & \left.
+ \frac{1}{2 m_W^2 s_w^2}
\left(
\frac{1}{2}
\sum_{i=1}^3 \left(V_{u_i d} V_{d u_i}^\dagger m_{u_i}^2 \right)
+ m_d^2 \left(\frac{1}{4} + I_{3d}^2\right)
\right)
\right] 
\eqa

\section{Tests and findings \label{sec:4}}
We performed several checks on our formulae. First of all, we 
derived them by means of two independent calculations, secondly, 
we explicitly checked the gauge invariance of our results with the help of
the Ward Identities listed in app.~\ref{appa}, that we derived, 
by using the Background Field Method described in~\cite{bfm1}, in the way 
we detail in the appendix. Given the fact
that only ${\rm R= R_1+R_2}$ is gauge invariant, 
we adopted the following strategy. The terms proportional
to $\lambda_{HV}$ in our effective vertices are expected to be gauge invariant
by themselves. Such terms can only be generated 
by ${\rm R_2}$, so that we could 
explicitly check, by using {\tt FORM}, that this part of our results
fulfills all of the Ward identities of app.~\ref{appa}, both in 
the 't Hooft-Feynman gauge and in the Background Field Method approach.
This provides an explicit test of the gauge invariance of the 
Four Dimensional Helicity regularization scheme in the complete Standard Model 
at 1-loop, and we consider this result as a by-product of our calculation.

To also test the parts not proportional to $\lambda_{HV}$, we 
computed analytically ${\rm R_1}$ 
\footnote{We extracted the ${\rm R_1}$ part
of the contributing tensor integrals by using the 
Passarino-Veltman~\cite{pasve} reduction technique and by further checking numerically the expressions with the 
help of CutTools~\cite{cuttools}.}, we added it to ${\rm R_2}$ and checked that
the quantity ${\rm R_1} + {\rm R_2}$  fulfills all of the 2-point and 3-point
Ward identities listed in the appendix. In the 4-point case, many new vertices 
are present in ${\rm R_1}$ that do not contribute to ${\rm R_2}$, such as VVVS, and, given 
the fact that, after all, we just need to check ${\rm R_2}$, we limited 
ourselves to verify the first six 4-point Ward identities given 
in app.~\ref{appa4},  which are the only ones including both the
VVVV and VVV vertices, but not VVVS. 
The described gauge invariance test on  ${\rm R_1+R_2}$ is a very powerful 
and non trivial one. In fact, 
the analytic expressions for ${\rm R_1}$ are, in general, much
more complicated than the ones for ${\rm R_2}$, involving a huge amount of 
terms with different combinations/powers of Gram determinants. 

\section{Conclusions \label{sec:5}}
In the last few years, new techniques have been developed to
efficiently deal with the problem of computing the 
radiative corrections needed to cope with the complicated phenomenology 
expected at LHC and ILC. Nowadays, thanks to the OPP technique, the 
so called Cut Constructible part  of the virtual 1-loop amplitudes can be obtained, in a purely numerical way, by means of a calculation of 
the same conceptual complexity of a tree level one.
 However, the determination of the remaining rational part ${\rm R}$ 
of the amplitude requires a different strategy. 
In the treatment at the {\em integrand level}, that we follow in this paper, a 
piece of ${\rm R}$, called ${\rm R_1}$, can be directly 
linked to the Cut Constructible part of the amplitude, and it is therefore 
numerically and automatically produced, in the OPP framework, by codes like CutTools. The remaining part of ${\rm R}$, called
${\rm R_2}$, cannot be determined numerically in 4 dimensions, and requires an
explicit computation in terms of the vertices of the theory at hand, 
up to four external legs. From the knowledge of these vertices, 
a finite set of effective tree level Feynman rules can be extracted to
be used to compute ${\rm R_2}$ for processes with an arbitrary 
number of external legs. Such effective ${\rm R_2}$ Feynman rules 
have been already given, in the literature, for QED and QCD and, 
in this paper, we completed the list by computing and presenting the set of ${\rm R_2}$ Feynman rules 
for the Electroweak sector, which was the last missing piece for
completely automatizing, in the framework of the OPP method, the 1-loop 
calculations in the $SU(3) \times SU(2) \times U(1)$ Standard Model.

In addition, since ${\rm R_2}$ is the only part of the amplitude sensitive to the choice of the regularization scheme, we explicitly proved, by checking  a large set of Ward identities, the gauge invariance of the 
Four Dimensional Helicity regularization  scheme in the full Electroweak 
sector at 1-loop.

\section*{Acknowledgments}
Many thanks to Huasheng Shao for having recomputed, independently of us, all
of the ${\rm R_2}$ effective vertices. Thanks to his help and by comparing
with an independent computation we performed in a general $R_\xi$ gauge,  
we could fix, in the present version, a few problems in our formulae.
We would like to thank Costas Papadopoulos, Ronald Kleiss and Stefan 
Dittmaier for useful discussions. R.P. is also indebted with Mauro Moretti
and Fulvio Piccinini for discussing the gauge invariance arguments presented 
in section~\ref{sec:2}.
R.P.'s and I.M.'s research was partially supported by the RTN
European Programme MRTN-CT-2006-035505 (HEPTOOLS, Tools and Precision
Calculations for Physics Discoveries at Colliders).
 M.V.G.'s research was supported by INFN.
The research of R.P. and M.V.G. was also supported 
by the MEC project FPA2008-02984.

\section*{Appendices}
\appendix
\section{Ward identities \label{appa}}
The Background Field Method (BFM) is a technique for quantizing gauge theories without losing explicit gauge invariance of the effective action~\cite{bfm1,bfm2,bfm3,bfm4,bfm5}. Starting from a classical Lagrangian, one can achieve this by decomposing the usual fields into background fields and quantum fields. Then, the background fields are treated as external sources, while the quantum fields are variables of integration in the functional integral. A gauge fixing term is added, which only breaks the invariance with respect to the quantum gauge transformations, while the invariance with respect to background-field gauge transformations is preserved.
From the Lagrangian mentioned above, one can construct an effective action $\Gamma[\hat{V},\hat{S},F,\bar{F}]$, where $\hat{V}$ refers to the background gauge fields, $\hat{S}$ to the background scalar fields and $F,\bar{F}$ to the fermion fields (for all fields that do not enter the gauge-fixing term, quantization is identical in the BFM and in the conventional formalism. Their Feynman rules for the background fields and quantum fields are also identical, so there is no need to distinguish them). This effective action is invariant under the background gauge transformations given in eqs. 21, 22 of~\cite{bfm1}. This invariance implies that 
\bqa
\frac{\delta\Gamma}{\delta \hat{\theta}^a}=0 \, ,
\eqa
where $a=A,Z,W^\pm$  and $\hat{\theta}^a$ are the infinitesimal gauge transformations of the background fields. By combining these formulas with eqs. 21, 22 of~\cite{bfm1}, one can produce eqs. 4, 5 and 6 of~\cite{bfm4}. By differentiating them with respect to background fields and setting the fields equal to zero, one obtains Ward identities for the vertex functions that are precisely the Ward identities related to the classical Lagrangian. In the papers ~\cite{bfm1} and ~\cite{bfm4} some of these Ward identities are listed (see also \cite{Bardin:1999ak}).  In the following, we extend this list 
by producing more Ward identities useful 
for our checks \footnote{We assume $V_{ud}= V^{\dagger}_{du}= 1$ and understand a sum over colors.}. 
\subsection{Ward identities involving VV, VS and SS}

\begin{equation}
k^{\mu}\Gamma^{AA}_{\mu\nu}(k,-k)=k^{\mu}\Gamma^{AZ}_{\mu\nu}(k,-k)= 0
\end{equation}

\bqa 
k^{\mu}\Gamma^{ZZ}_{\mu\nu}(k,-k)-iM_Z \Gamma^{\chi Z}_{\nu}(k,-k)=0
\eqa

\bqa 
k^{\mu}\Gamma^{W^\pm W^\mp}_{\mu\nu}(k,-k)\mp M_W \Gamma^{\phi^\pm W^\mp}_{\nu}(k,-k)=0
\eqa

\bqa 
k^{\mu}\Gamma^{Z\chi}_{\mu\nu}(k,-k)-iM_Z \Gamma^{\chi \chi}_{\nu}(k,-k)
+\frac{ie}{2 c_w s_w} T^H = 0
\eqa

\bqa 
k^{\mu}\Gamma^{W^\pm \phi^\mp}_{\mu\nu}(k,-k)\mp M_W \Gamma^{\phi^\pm \phi^\mp}_{\nu}(k,-k) \pm \frac{e}{2 s_w} T^H = 0 \, 
\eqa
In the previous identities, $T^H$ is the Higgs tadpole contribution. We have 
found a non-vanishing ${\rm R_2}$ contribution to $T^H$, due to the coupling 
of $H$ with $Z$ and $W$, while ${\rm R_1}$ does not contribute to $T^H$.

\subsection{Ward identities involving VFF, SFF and FF}

\bqa 
k^{\mu}\Gamma^{A \bar{f} f}_{\mu}(k,\bar{p},p) +e Q_f(\Gamma^{\bar{f} f}(\bar{p},k+p)-\Gamma^{\bar{f} f}(k+\bar{p},p))=0
\eqa

\bqa 
k^{\mu}\Gamma^{Z \bar{f} f}_{\mu}(k,\bar{p},p) -iM_Z\Gamma^{\chi \bar{f} f}(k,\bar{p},p)
-e (\Gamma^{\bar{f} f}(\bar{p},k+p)(v_f-a_f\gamma_5)\nonumber \\
-(v_f+a_f\gamma_5)\Gamma^{\bar{f} f}(k+\bar{p},p))=0 \, 
\eqa

\bqa 
k^{\mu}\Gamma^{W^+ \bar{f_u} f_d}_{\mu}(k,\bar{p},p) -M_W\Gamma^{\phi^+ \bar{f_u} f_d}(k,\bar{p},p)
-\frac{e}{\sqrt{2} s_w} (\Gamma^{\bar{f_u} f_u}(\bar{p},k+p)\Omega_- 
- \nonumber \\
\Omega_+\Gamma^{\bar{f_d} f_d}(k+\bar{p},p))=0
\eqa

\bqa 
k^{\mu}\Gamma^{W^- \bar{f_d} f_u}_{\mu}(k,\bar{p},p) +M_W\Gamma^{\phi^- \bar{f_d} f_u}(k,\bar{p},p)
-\frac{e}{\sqrt{2} s_w} (\Gamma^{\bar{f_d} f_d}(\bar{p},k+p)\Omega_- - \nonumber \\
\Omega_+\Gamma^{\bar{f_u} f_u}(k+\bar{p},p))=0
\eqa
In the previous expressions, $f_u$ is a fermion with $I_{3f}$=1/2, 
$f_d$ is the fermion of the same weak-isospin doublet with $I_{3f}$= -1/2, 
$v_f = (I_{3f} - 2 s_w^2 Q_f)/(2 s_w c_w)$ and $a_f = I_{3f}/(2 s_w c_w)$.

\subsection{Ward identities involving VVV, VVS and VV}

\bqa
k^{\mu}\Gamma^{AW^+W^-}_{\mu\nu\sigma}(k,k_+,k_-)  -e(\Gamma^{W^+W^-}_{\nu\sigma}(k_+,k+k_-) 
-\Gamma^{W^+W^-}_{\nu\sigma}(k+k_+,k_-))=0
\eqa

\bqa
k_+^{\mu}\Gamma^{W^+W^-A}_{\mu\nu\sigma}(k_+,k_-,k) -M_W \Gamma^{\phi^+W^-A}_{\nu\sigma}(k_+,k_-,k) -e\Gamma^{W^+W^-}_{\sigma\nu}(k+k_+,k_-) \nonumber \\
+e(\Gamma^{AA}_{\sigma\nu}(k,k_++k_-)-\frac{c_w}{s_w}\Gamma^{AZ}_{\sigma\nu}(k,k_++k_-))=0
\eqa

\bqa
k_-^{\mu}\Gamma^{W^-W^+A}_{\mu\nu\sigma}(k_-,k_+,k) +M_W \Gamma^{\phi^-W^+A}_{\nu\sigma}(k_-,k_+,k) +e\Gamma^{W^-W^+}_{\sigma\nu}(k+k_-,k_+) \nonumber \\
 -e(\Gamma^{AA}_{\sigma\nu}(k,k_++k_-)-\frac{c_w}{s_w}\Gamma^{AZ}_{\sigma\nu}(k,k_++k_-))=0
\eqa

\bqa
k^{\mu}\Gamma^{ZW^+W^-}_{\mu\nu\sigma}(k,k_+,k_-) -iM_Z \Gamma^{\chi W^+W^-}_{\nu\sigma}(k,k_+,k_-) -e\frac{c_w}{s_w}(\Gamma^{W^+W^-}_{\nu\sigma}(k+k_+,k_-) \nonumber \\
-\Gamma^{W^-W^+}_{\sigma\nu}(k+k_-,k_+))=0
\eqa

\bqa
k_+^{\mu}\Gamma^{W^+W^-Z}_{\mu\nu\sigma}(k_+,k_-,k) -M_W \Gamma^{\phi^+W^-Z}_{\nu\sigma}(k_+,k_-,k) +e\frac{c_w}{s_w}\Gamma^{W^+W^-}_{\sigma\nu}(k+k_+,k_-) \nonumber \\
+e(\Gamma^{AZ}_{\nu\sigma}(k_++k_-,k)-\frac{c_w}{s_w}\Gamma^{ZZ}_{\nu\sigma}(k_++k_-,k))=0
\eqa

\bqa
k_-^{\mu}\Gamma^{W^-W^+Z}_{\mu\nu\sigma}(k_-,k_+,k) +M_W \Gamma^{\phi^-W^+Z}_{\nu\sigma}(k_-,k_+,k) -e\frac{c_w}{s_w}\Gamma^{W^-W^+}_{\sigma\nu}(k+k_-,k_+) \nonumber \\
 -e(\Gamma^{AZ}_{\nu\sigma}(k_++k_-,k)-\frac{c_w}{s_w}\Gamma^{ZZ}_{\nu\sigma}(k_++k_-,k))=0
\eqa

\subsection{Ward identities involving VVS, VSS and VS}

\bqa
k_1^{\mu}\Gamma^{AAH}_{\mu\nu}(k_1,k_2,k_3)=k_1^{\mu}\Gamma^{AA\chi}_{\mu\nu}(k_1,k_2,k_3)\nonumber \\
=k_1^{\mu}\Gamma^{AZH}_{\mu\nu}(k_1,k_2,k_3)=k_1^{\mu}\Gamma^{AZ\chi}_{\mu\nu}(k_1,k_2,k_3)=0
\eqa

\bqa
k^{\mu}\Gamma^{AW^+\phi^-}_{\mu\nu}(k,k_+,k_-)+ e\Gamma^{W^+\phi^-}_{\nu}(k+k_+,k_-)-e\Gamma^{\phi^-W^+}_{\nu}(k+k_-,k_+)=0
\eqa

\bqa
k^{\mu}\Gamma^{AW^-\phi^+}_{\mu\nu}(k,k_-,k_+)- e\Gamma^{W^-\phi^+}_{\nu}(k+k_-,k_+)+e\Gamma^{\phi^+W^-}_{\nu}(k+k_+,k_-)=0
\eqa

\bqa
k_1^{\mu}\Gamma^{ZAH}_{\mu\nu}(k_1,k_2,k_3) -iM_Z\Gamma^{\chi AH}_{\nu}(k_1,k_2,k_3)-\frac{ie}{2c_ws_w}\Gamma^{\chi A}_{\nu}(k_1+k_3,k_2)=0
\eqa

\bqa
k_1^{\mu}\Gamma^{ZA\chi}_{\mu\nu}(k_1,k_2,k_3) -iM_Z\Gamma^{\chi A\chi}_{\nu}(k_1,k_2,k_3)+\frac{ie}{2c_ws_w}\Gamma^{HA}_{\nu}(k_1+k_3,k_2)=0
\eqa

\bqa
k_1^{\mu}\Gamma^{ZZH}_{\mu\nu}(k_1,k_2,k_3) -iM_Z\Gamma^{\chi ZH}_{\nu}(k_1,k_2,k_3)-\frac{ie}{2c_ws_w}\Gamma^{\chi Z}_{\nu}(k_1+k_3,k_2)=0
\eqa

\bqa
k_1^{\mu}\Gamma^{ZZ\chi}_{\mu\nu}(k_1,k_2,k_3) -iM_Z\Gamma^{\chi Z\chi}_{\nu}(k_1,k_2,k_3)+\frac{ie}{2c_ws_w}\Gamma^{HZ}_{\nu}(k_1+k_3,k_2)=0
\eqa

\bqa
k^{\mu}\Gamma^{ZW^+\phi^-}_{\mu\nu}(k,k_+,k_-) -iM_Z\Gamma^{\chi W^+\phi^-}_{\nu}(k,k_+,k_-)\nonumber\\
-e\frac{c_w}{s_w}\Gamma^{W^+\phi^-}_{\nu}(k+k_+,k_-)+e\frac{c_w^2-s_w^2}{2c_ws_w}\Gamma^{\phi^-W^+}_{\nu}(k+k_-,k_+)=0
\eqa

\bqa
k^{\mu}\Gamma^{ZW^-\phi^+}_{\mu\nu}(k,k_-,k_+) -iM_Z\Gamma^{\chi W^-\phi^+}_{\nu}(k,k_-,k_+)\nonumber\\
+e\frac{c_w}{s_w}\Gamma^{W^-\phi^+}_{\nu}(k+k_-,k_+)-e\frac{c_w^2-s_w^2}{2c_ws_w}\Gamma^{\phi^+W^-}_{\nu}(k+k_+,k_-)=0
\eqa

\bqa
k_+^{\mu}\Gamma^{W^+A\phi^-}_{\mu\nu}(k_+,k,k_-) -M_W\Gamma^{\phi^+A\phi^-}_{\nu}(k_+,k,k_-)\nonumber\\
-e\Gamma^{W^+\phi^-}_{\nu}(k+k_+,k_-)+\frac{e}{2s_w}(\Gamma^{HA}_{\nu}(k_++k_-,k)+i\Gamma^{\chi A}_{\nu}(k_++k_-,k))=0
\eqa

\bqa
k_+^{\mu}\Gamma^{W^+Z\phi^-}_{\mu\nu}(k_+,k,k_-) -M_W\Gamma^{\phi^+Z\phi^-}_{\nu}(k_+,k,k_-)\nonumber\\
+e\frac{c_w}{s_w}\Gamma^{W^+\phi^-}_{\nu}(k+k_+,k_-)+\frac{e}{2s_w}(\Gamma^{HZ}_{\nu}(k_++k_-,k)+i\Gamma^{\chi Z}_{\nu}(k_++k_-,k))=0
\eqa

\bqa
k_+^{\mu}\Gamma^{W^+W^-H}_{\mu\nu}(k_+,k_-,k) -M_W\Gamma^{\phi^+W^-H}_{\nu}(k_+,k_-,k)\nonumber\\
-\frac{e}{2s_w}\Gamma^{\phi^+W^-}_{\nu}(k+k_+,k_-)+e(\Gamma^{AH}_{\nu}(k_++k_-,k)-\frac{c_w}{s_w}\Gamma^{ZH}_{\nu}(k_++k_-,k))=0
\eqa

\bqa
k_+^{\mu}\Gamma^{W^+W^-\chi}_{\mu\nu}(k_+,k_-,k) -M_W\Gamma^{\phi^+W^-\chi}_{\nu}(k_+,k_-,k)\nonumber\\
-\frac{ie}{2s_w}\Gamma^{\phi^+W^-}_{\nu}(k+k_+,k_-)+e(\Gamma^{A\chi}_{\nu}(k_++k_-,k)-\frac{c_w}{s_w}\Gamma^{Z\chi}_{\nu}(k_++k_-,k))=0
\eqa

\bqa
k_-^{\mu}\Gamma^{W^-A\phi^+}_{\mu\nu}(k_-,k,k_+) +M_W\Gamma^{\phi^-A\phi^+}_{\nu}(k_-,k,k_+)\nonumber\\
+e\Gamma^{W^-\phi^+}_{\nu}(k+k_-,k_+)-\frac{e}{2s_w}(\Gamma^{HA}_{\nu}(k_++k_-,k)-i\Gamma^{\chi A}_{\nu}(k_++k_-,k))=0
\eqa

\bqa
k_-^{\mu}\Gamma^{W^-Z\phi^+}_{\mu\nu}(k_-,k,k_+) +M_W\Gamma^{\phi^-Z\phi^+}_{\nu}(k_-,k,k_+)\nonumber\\
-e\frac{c_w}{s_w}\Gamma^{W^-\phi^+}_{\nu}(k+k_-,k_+)-\frac{e}{2s_w}(\Gamma^{HZ}_{\nu}(k_++k_-,k)-i\Gamma^{\chi Z}_{\nu}(k_++k_-,k))=0
\eqa

\bqa
k_-^{\mu}\Gamma^{W^-W^+H}_{\mu\nu}(k_-,k_+,k) +M_W\Gamma^{\phi^-W^+H}_{\nu}(k_-,k_+,k)\nonumber\\
+\frac{e}{2s_w}\Gamma^{\phi^-W^+}_{\nu}(k+k_-,k_+)-e(\Gamma^{AH}_{\nu}(k_++k_-,k)-\frac{c_w}{s_w}\Gamma^{ZH}_{\nu}(k_++k_-,k))=0
\eqa

\bqa
k_-^{\mu}\Gamma^{W^-W^+\chi}_{\mu\nu}(k_-,k_+,k) +M_W\Gamma^{\phi^-W^+\chi}_{\nu}(k_-,k_+,k)\nonumber\\
-\frac{ie}{2s_w}\Gamma^{\phi^-W^+}_{\nu}(k+k_-,k_+)-e(\Gamma^{A\chi}_{\nu}(k_++k_-,k)-\frac{c_w}{s_w}\Gamma^{Z\chi}_{\nu}(k_++k_-,k))=0
\eqa


\subsection{Ward identities involving VSS, SSS and SS}

\bqa
k_1^{\mu}\Gamma^{AHH}_{\mu}(k_1,k_2,k_3)=k_1^{\mu}\Gamma^{AH\chi}_{\mu}(k_1,k_2,k_3)=k_1^{\mu}\Gamma^{A\chi\chi}_{\mu}(k_1,k_2,k_3)=0
\eqa

\bqa
k^{\mu}\Gamma^{A\phi^+ \phi^-}_{\mu}(k,k_+,k_-) +e(\Gamma^{\phi^+ \phi^-}(k+k_+,k_-)-\Gamma^{\phi^-\phi^+}(k+k_-,k_+))=0
\eqa

\bqa
 k_1^{\mu}\Gamma^{ZHH}_{\mu}(k_1,k_2,k_3) -iM_Z\Gamma^{\chi HH}(k_1,k_2,k_3)
& - & \frac{ie}{2c_ws_w}   \left( \Gamma^{\chi H}(k_1+k_2,k_3) \right.\nonumber\\
&+& \left.\Gamma^{\chi H}(k_1+k_3,k_2)\right)=0
\eqa

\bqa
k_1^{\mu}\Gamma^{ZH\chi}_{\mu}(k_1,k_2,k_3) -iM_Z\Gamma^{\chi H\chi}(k_1,k_2,k_3)
& - & \frac{ie}{2c_ws_w}\left(\Gamma^{\chi\chi}(k_1+k_2,k_3) \right. \nonumber\\
&-& \left.\Gamma^{HH}(k_1+k_3,k_2)\right)=0
\eqa

\bqa
k_1^{\mu}\Gamma^{Z\chi \chi }_{\mu}(k_1,k_2,k_3) -iM_Z\Gamma^{\chi \chi \chi }(k_1,k_2,k_3)
& + & \frac{ie}{2c_ws_w}\left(\Gamma^{H\chi }(k_1+k_2,k_3)\right.\nonumber\\
&+&\left.\Gamma^{H\chi}(k_1+k_3,k_2)\right)=0
\eqa

\bqa
k^{\mu}\Gamma^{Z\phi^+\phi^-}_{\mu}(k,k_+,k_-) -iM_Z\Gamma^{\chi\phi^+\phi^-}(k,k_+,k_-)
\nonumber\\
-e\frac{c_w^2-s_w^2}{2c_ws_w}(\Gamma^{\phi^+\phi^-}(k+k_+,k_-)-\Gamma^{\phi^-\phi^+}(k+k_-,k_+))=0
\eqa

\bqa
k_+^{\mu}\Gamma^{W^+H\phi^-}_{\mu}(k_+,k,k_-) -M_W\Gamma^{\phi^+H\phi^-}(k_+,k,k_-)
\nonumber\\
+\frac{e}{2s_w}(\Gamma^{HH}(k_-+k_+,k)+i\Gamma^{\chi H}(k_++k_-,k))
-\frac{e}{2s_w}\Gamma^{\phi^+\phi^-}(k_++k,k_-)=0
\eqa

\bqa
k_+^{\mu}\Gamma^{W^+\chi\phi^-}_{\mu}(k_+,k,k_-) -M_W\Gamma^{\phi^+\chi\phi^-}(k_+,k,k_-)
\nonumber\\
+\frac{e}{2s_w}(\Gamma^{H\chi}(k_-+k_+,k)+i\Gamma^{\chi\chi}(k_++k_-,k))
-\frac{ie}{2s_w}\Gamma^{\phi^+\phi^-}(k_++k,k_-)=0
\eqa

\bqa
k_-^{\mu}\Gamma^{W^-H\phi^+}_{\mu}(k_-,k,k_+) +M_W\Gamma^{\phi^-H\phi^+}(k_-,k,k_+)\nonumber\\
-\frac{e}{2s_w}(\Gamma^{HH}(k_-+k_+,k)-i\Gamma^{\chi H}(k_++k_-,k))+\frac{e}{2s_w}\Gamma^{\phi^-\phi^+}(k_-+k,k_+)=0
\eqa

\bqa
k_-^{\mu}\Gamma^{W^-\chi\phi^+}_{\mu}(k_-,k,k_+) +M_W\Gamma^{\phi^-\chi\phi^+}(k_-,k,k_+)
\nonumber\\
-\frac{e}{2s_w}(\Gamma^{H\chi}(k_-+k_+,k)-i\Gamma^{\chi\chi}(k_++k_-,k))-\frac{ie}{2s_w}\Gamma^{\phi^-\phi^+}(k_-+k,k_+)=0
\eqa

\subsection{\label{appa4} Ward identities involving VVVV, VVVS and VVV}

\bqa
k_{1,2,3,4}^{\mu}\Gamma^{AAAA}_{\mu\nu\kappa\sigma}(k_1,k_2,k_3,k_4) =0
\eqa

\bqa
k_{1,2,3}^{\mu}\Gamma^{AAAZ}_{\mu\nu\kappa\sigma}(k_1,k_2,k_3,k_4) =0
\eqa

\bqa
k_{1,2}^{\mu}\Gamma^{AAZZ}_{\mu\nu\kappa\sigma}(k_1,k_2,k_3,k_4) =0
\eqa

\bqa
k_1^{\mu}\Gamma^{AZZZ}_{\mu\nu\kappa\sigma}(k_1,k_2,k_3,k_4) =0
\eqa

\bqa
k_1^{\mu}\Gamma^{AAW^+W^-}_{\mu\nu\kappa\sigma}(k_1,k_2,k_+,k_-) 
+ & e & \left[  
\Gamma^{AW^+W^-}_{\nu\kappa\sigma}(k_2,k_1+k_+,k_-) \right.\nonumber\\
&-& \left. \Gamma^{AW^+W^-}_{\nu\kappa\sigma}(k_2,k_+,k_1+k_-) \right]=0 
\eqa

\bqa
k_1^{\mu}\Gamma^{AZW^+W^-}_{\mu\nu\kappa\sigma}(k_1,k_2,k_+,k_-) +&e& \left[\Gamma^{W^+ZW^-}_{\kappa\nu\sigma}(k_1+k_+,k_2,k_-) \right.\nonumber\\
&-&\left.\Gamma^{W^-ZW^+}_{\sigma\nu\kappa}(k_1+k_-,k_2,k_+) \right]=0 
\eqa

\bqa
k^{\mu}\Gamma^{ZV_2V_3V_4}_{\mu\nu\kappa\sigma}(k_1,k_2,k_3,k_4)-i M_Z \Gamma^{\chi V_2V_3V_4}_{\nu\kappa\sigma}(k_1,k_2,k_3,k_4)=0 \, ,
\eqa
where $k$ here refers to any of the Z momenta and V's stand for A or Z.

\bqa
& k_+^{\mu} &\Gamma^{W^+W^-AA}_{\mu\nu\kappa\sigma}(k_+,k_-,k_3,k_4) +e\left[\Gamma^{AAA}_{\nu\kappa\sigma}(k_++k_-,k_3,k_4)
-\frac{c_w}{s_w}\Gamma^{ZAA}_{\nu\kappa\sigma}(k_++k_-,k_3,k_4) \right. \nonumber \\
&-&  \left.\Gamma^{W^+W^-A}_{\sigma\nu\kappa}(k_++k_4,k_-,k_3) -
 \Gamma^{W^-W^+A}_{\nu\kappa\sigma}(k_-,k_++k_3,k_4) \right]\nonumber\\
&-& M_W \Gamma^{\phi^+W^-AA}_{\nu\kappa\sigma}(k_+,k_-,k_3,k_4)=0 
\eqa

\bqa
&k_-^{\mu}&\Gamma^{W^-W^+AA}_{\mu\nu\kappa\sigma}(k_-,k_+,k_3,k_4) -e\left[\Gamma^{AAA}_{\nu\kappa\sigma}(k_++k_-,k_3,k_4)
-\frac{c_w}{s_w}\Gamma^{ZAA}_{\nu\kappa\sigma}(k_++k_-,k_3,k_4) 
\right.\nonumber \\
&-&  \left.\Gamma^{W^-W^+A}_{\sigma\nu\kappa}(k_-+k_4,k_+,k_3)
-\Gamma^{W^+W^-A}_{\nu\kappa\sigma}(k_+,k_-+k_3,k_4) \right]\nonumber\\
&+& M_W \Gamma^{\phi^-W^+AA}_{\nu\kappa\sigma}(k_-,k_+,k_3,k_4)=0
\eqa

\bqa
& k_1^{\mu} & \Gamma^{ZZW^+W^-}_{\mu\nu\kappa\sigma}(k_1,k_2,k_+,k_-) -e\frac{c_w}{s_w}\left[\Gamma^{W^+ZW^-}_{\kappa\nu\sigma}(k_1+k_+,k_2,k_-) \right.
\nonumber\\
&-&\left.\Gamma^{W^-ZW^+}_{\sigma\nu\kappa}(k_1+k_-,k_2,k_+) \right]  
 -  iM_Z \Gamma^{\chi ZW^+W^-}_{\nu\kappa\sigma}(k_1,k_2,k_+,k_-)=0
\eqa

\bqa
&k_+^{\mu}&\Gamma^{W^+ZZW^-}_{\mu\nu\kappa\sigma}(k_+,k_1,k_2,k_-) +e\left[\Gamma^{AZZ}_{\sigma\nu\kappa}(k_++k_-,k_1,k_2)
-\frac{c_w}{s_w}\Gamma^{ZZZ}_{\sigma\nu\kappa}(k_++k_-,k_1,k_2)\right.\nonumber\\
& + & \left. \frac{c_w}{s_w}\Gamma^{W^+ZW^-}_{\nu\kappa\sigma}(k_++k_1,k_2,k_-) 
+ \frac{c_w}{s_w}\Gamma^{W^+ZW^-}_{\kappa\nu\sigma}(k_++k_2,k_1,k_-) \right]
\nonumber\\
&-& M_W \Gamma^{\phi^+ZZW^-}_{\nu\kappa\sigma}(k_+,k_1,k_2,k_-)
=0  
\eqa

\bqa
&k_-^{\mu}&\Gamma^{W^-W^+ZZ}_{\mu\nu\kappa\sigma}(k_-,k_+,k_3,k_4) -e\left[\Gamma^{AZZ}_{\nu\kappa\sigma}(k_++k_-,k_3,k_4)
-\frac{c_w}{s_w}\Gamma^{ZZZ}_{\nu\kappa\sigma}(k_++k_-,k_3,k_4)\right.\nonumber\\
&+& \left. \frac{c_w}{s_w}\Gamma^{W^-W^+Z}_{\kappa\nu\sigma}(k_-+k_3,k_+,k_4)
+ \frac{c_w}{s_w}\Gamma^{W^-W^+Z}_{\sigma\nu\kappa}(k_-+k_4,k_+,k_3)\right]
\nonumber\\
&+& M_W \Gamma^{\phi^-W^+ZZ}_{\nu\kappa\sigma}(k_-,k_+,k_3,k_4)
=0
\eqa

\bqa
& k_1^{\mu}& \Gamma^{ZAW^+W^-}_{\mu\nu\kappa\sigma}(k_1,k_2,k_+,k_-) -e\frac{c_w}{s_w}\left[\Gamma^{W^+AW^-}_{\kappa\nu\sigma}(k_1+k_+,k_2,k_-) \right.\nonumber \\
 & - & \left. \Gamma^{W^-AW^+}_{\sigma\nu\kappa}(k_1+k_-,k_2,k_+) \right] 
 -  iM_Z \Gamma^{\chi AW^+W^-}_{\nu\kappa\sigma}(k_1,k_2,k_+,k_-)=0
\eqa

\bqa
& k_+^{\mu} &\Gamma^{W^+W^-AZ}_{\mu\nu\kappa\sigma}(k_+,k_-,k_3,k_4) 
+e \left[\Gamma^{AAZ}_{\nu\kappa\sigma}(k_++k_-,k_3,k_4)
-\frac{c_w}{s_w}\Gamma^{ZAZ}_{\nu\kappa\sigma}(k_++k_-,k_3,k_4) \right. 
\nonumber \\
& - & \left.\Gamma^{W^+W^-Z}_{\kappa\nu\sigma}(k_++k_3,k_-,k_4) 
+\frac{c_w}{s_w}\Gamma^{W^+W^-A}_{\sigma\nu\kappa}(k_++k_4,k_-,k_3)\right]
\nonumber \\
& - & M_W \Gamma^{\phi^+W^-AZ}_{\nu\kappa\sigma}(k_+,k_-,k_3,k_4) =0 
\eqa

\bqa
& k_-^{\mu} & \Gamma^{W^-W^+AZ}_{\mu\nu\kappa\sigma}(k_-,k_+,k_3,k_4) -e \left[\Gamma^{AAZ}_{\nu\kappa\sigma}(k_++k_-,k_3,k_4)
-\frac{c_w}{s_w}\Gamma^{ZAZ}_{\nu\kappa\sigma}(k_++k_-,k_3,k_4) \right. \nonumber \\
& - &  \left.\Gamma^{W^-W^+Z}_{\kappa\nu\sigma}(k_-+k_3,k_+,k_4) 
+\frac{c_w}{s_w}\Gamma^{W^-W^+A}_{\sigma\nu\kappa}(k_-+k_4,k_+,k_3)
\right]\nonumber \\
&+& M_W \Gamma^{\phi^-W^+AZ}_{\nu\kappa\sigma}(k_-,k_+,k_3,k_4) = 0 
\eqa

\bqa
&k_{1+}^{\mu}&\Gamma^{W^+W^-W^+W^-}_{\mu\nu\kappa\sigma}(k_{1+},k_{1-},k_{2+},k_{2-}) +e\left[\Gamma^{AW^+W^-}_{\nu\kappa\sigma}(k_{1+}+k_{1-},k_{2+},k_{2-})\right.
\nonumber\\ 
&-&\left.\frac{c_w}{s_w}\Gamma^{ZW^+W^-}_{\nu\kappa\sigma}(k_{1+}+k_{1-},k_{2+},k_{2-})
+\Gamma^{AW^-W^+}_{\sigma\nu\kappa}(k_{1+}+k_{2-},k_{1-},k_{2+})
\right.\nonumber\\
&-&\left.\frac{c_w}{s_w}\Gamma^{ZW^-W^+}_{\sigma\nu\kappa}(k_{1+}+k_{2-},k_{1-},k_{2+})\right]- M_W \Gamma^{\phi^+W^-W^+W^-}_{\nu\kappa\sigma}(k_{1+},k_{1-},k_{2+},k_{2-})=0
\nonumber\\
\eqa

\bqa
&k_{1-}^{\mu}&\Gamma^{W^-W^+W^-W^+}_{\mu\nu\kappa\sigma}(k_{1-},k_{1+},k_{2-},k_{2+}) -e\left[\Gamma^{AW^-W^+}_{\nu\kappa\sigma}(k_{1+}+k_{1-},k_{2-},k_{2+})\right.
\nonumber\\
&-&\left.
\frac{c_w}{s_w}\Gamma^{ZW^-W^+}_{\nu\kappa\sigma}(k_{1+}+k_{1-},k_{2-},k_{2+}) 
+\Gamma^{AW^+W^-}_{\sigma\nu\kappa}(k_{2+}+k_{1-},k_{1+},k_{2-})\right.
\nonumber\\
&-&\left.
\frac{c_w}{s_w}\Gamma^{ZW^+W^-}_{\sigma\nu\kappa}(k_{2+}+k_{1-},k_{1+},k_{2-})
\right]
+ M_W \Gamma^{\phi^-W^+W^-W^+}_{\nu\kappa\sigma}(k_{1-},k_{1+},k_{2-},k_{2+})=0
\nonumber\\
\eqa

\subsection{Ward identities involving SSSS, VSSS and SSS}

\bqa
&k_1^{\mu}&\Gamma^{Z\chi HH}_{\mu}(k_1,k_2,k_3,k_4) -iM_Z\Gamma^{\chi \chi HH}(k_1,k_2,k_3,k_4)
-\frac{ie}{2c_ws_w}\left[\Gamma^{\chi \chi H}(k_1+k_3,k_2,k_4)\right.\nonumber\\
 &+& \left.\Gamma^{\chi \chi H}(k_1+k_4,k_2,k_3)-\Gamma^{HHH}(k_1+k_2,k_3,k_4)
\right]=0
\eqa

\bqa
&k_1^{\mu}&\Gamma^{Z\chi \chi \chi }_{\mu}(k_1,k_2,k_3,k_4) -iM_Z\Gamma^{\chi \chi \chi \chi }(k_1,k_2,k_3,k_4)+\frac{ie}{2c_ws_w}\left[\Gamma^{H\chi \chi }(k_1+k_2,k_3,k_4) \right.\nonumber\\
&+&\left.\Gamma^{H\chi \chi }(k_1+k_3,k_2,k_4)+\Gamma^{H\chi \chi }(k_1+k_4,k_2,k_3)\right]=0
\eqa

\bqa
& k_1^{\mu} & \Gamma^{ZH\phi^+\phi^-}_{\mu}(k_1,k_2,k_+,k_-) -iM_Z\Gamma^{\chi H\phi^+\phi^-}(k_1,k_2,k_+,k_-) 
\nonumber \\
& - & e\frac{c_w^2-s_w^2}{2c_ws_w}\left[\Gamma^{\phi^+H\phi^-}(k_1+k_+,k_2,k_-)-\Gamma^{\phi^-H\phi^+}(k_1+k_-,k_2,k_+)\right] \nonumber\\
&-&\frac{ie}{2c_ws_w}\Gamma^{\chi \phi^+\phi^-}(k_1+k_2,k_+,k_-) =0
\eqa

\bqa
& k_1^{\mu} & \Gamma^{Z\chi \phi^+\phi^-}_{\mu}(k_1,k_2,k_+,k_-) -iM_Z\Gamma^{\chi \chi \phi^+\phi^-}(k_1,k_2,k_+,k_-) \nonumber \\
& - & e\frac{c_w^2-s_w^2}{2c_ws_w}\left[\Gamma^{\phi^+\chi \phi^-}(k_1+k_+,k_2,k_-)-\Gamma^{\phi^-\chi \phi^+}(k_1+k_-,k_2,k_+)\right] \nonumber\\
&+&\frac{ie}{2c_ws_w}\Gamma^{H\phi^+\phi^-}(k_1+k_2,k_+,k_-)=0
\eqa

\bqa
&k_+^{\mu}&\Gamma^{W^+\phi^-HH}_{\mu}(k_+,k_-,k_1,k_2) -M_W\Gamma^{\phi^+\phi^-HH}(k_+,k_-,k_1,k_2)
\nonumber\\
& + &\frac{e}{2s_w}
\left[\Gamma^{HHH}(k_++k_-,k_1,k_2) + i\Gamma^{\chi HH}(k_++k_-,k_1,k_2) \right.
\nonumber\\
& - & \left. \Gamma^{\phi^+\phi^-H}(k_1+k_+,k_-,k_2)-\Gamma^{\phi^+\phi^-H}(k_2+k_+,k_-,k_1)\right]=0
\eqa

\bqa
& k_+^{\mu} & \Gamma^{W^+\phi^-H\chi }_{\mu}(k_+,k_-,k_1,k_2) -M_W\Gamma^{\phi^+\phi^-H\chi }(k_+,k_-,k_1,k_2) + \frac{e}{2s_w}\left[\Gamma^{HH\chi }(k_++k_-,k_1,k_2)
\right. \nonumber \\
& + & \left.
i\Gamma^{\chi H\chi }(k_++k_-,k_1,k_2)-\Gamma^{\phi^+\phi^-\chi }(k_1+k_+,k_-,k_2) 
- i \Gamma^{\phi^+\phi^-H}(k_2+k_+,k_-,k_1)\right]=0
\nonumber \\
\eqa

\bqa
&k_+^{\mu}&\Gamma^{W^+\phi^-\chi \chi }_{\mu}(k_+,k_-,k_1,k_2) -M_W\Gamma^{\phi^+\phi^-\chi \chi }(k_+,k_-,k_1,k_2)\nonumber\\
& + & \frac{e}{2s_w}\left[\Gamma^{H\chi \chi }(k_++k_-,k_1,k_2)
+ i\Gamma^{\chi \chi \chi }(k_++k_-,k_1,k_2)\right. 
\nonumber\\
& - & i \left.\Gamma^{\phi^+\phi^-\chi }(k_1+k_+,k_-,k_2) - i \Gamma^{\phi^+\phi^-\chi }(k_2+k_+,k_-,k_1)\right]=0
\eqa

\bqa
&k_{1+}^{\mu}&\Gamma^{W^+\phi^-\phi^+\phi^-}_{\mu}(k_{1+},k_{1-},k_{2+},k_{2-}) -M_W\Gamma^{\phi^+\phi^-\phi^+\phi^-}(k_{1+},k_{1-},k_{2+},k_{2-})\nonumber\\
&+&\frac{e}{2s_w}\left[\Gamma^{H\phi^+\phi^-}(k_{1+}+k_{1-},k_{2+},k_{2-})+i\Gamma^{\chi \phi^+\phi^-}(k_{1+}+k_{1-},k_{2+},k_{2-})\right.\nonumber\\
&+&\left.\Gamma^{H\phi^-\phi^+}(k_{1+}+k_{2-},k_{1-},k_{2+})+i\Gamma^{\chi \phi^-\phi^+}(k_{1+}+k_{2-},k_{1-},k_{2+})\right]=0
\eqa

\bqa
&k_-^{\mu}&\Gamma^{W^-\phi^+HH}_{\mu}(k_-,k_+,k_1,k_2) +M_W\Gamma^{\phi^-\phi^+HH}(k_-,k_+,k_1,k_2) \nonumber\\
& -& \frac{e}{2s_w}\left[\Gamma^{HHH}(k_++k_-,k_1,k_2)
- i\Gamma^{\chi HH}(k_++k_-,k_1,k_2)\right. \nonumber\\
 &-& \left. \Gamma^{\phi^-\phi^+H}(k_1+k_-,k_+,k_2) - \Gamma^{\phi^-\phi^+H}(k_2+k_-,k_+,k_1)\right]=0 
\eqa

\bqa
& k_-^{\mu} & \Gamma^{W^-\phi^+H\chi }_{\mu}(k_-,k_+,k_1,k_2) +M_W\Gamma^{\phi^-\phi^+H\chi }(k_-,k_+,k_1,k_2) \nonumber \\
&-&\frac{e}{2s_w}\left[\Gamma^{HH\chi }(k_++k_-,k_1,k_2)
 -   i\Gamma^{\chi H\chi }(k_++k_-,k_1,k_2)\right. \nonumber\\
&-& \left. \Gamma^{\phi^-\phi^+\chi }(k_1+k_-,k_+,k_2) + i \Gamma^{\phi^-\phi^+H}(k_2+k_-,k_+,k_1)\right]=0
\eqa

\bqa
&k_-^{\mu}&\Gamma^{W^-\phi^+\chi \chi }_{\mu}(k_-,k_+,k_1,k_2) +M_W\Gamma^{\phi^-\phi^+\chi \chi }(k_-,k_+,k_1,k_2) \nonumber\\
&-&\frac{e}{2s_w}\left[\Gamma^{H\chi \chi }(k_++k_-,k_1,k_2)
- i\Gamma^{\chi \chi \chi }(k_++k_-,k_1,k_2)\right. \nonumber\\
&+& i \left.\Gamma^{\phi^-\phi^+\chi }(k_1+k_-,k_+,k_2) + i \Gamma^{\phi^-\phi^+\chi }(k_2+k_-,k_+,k_1)\right]=0
\eqa

\bqa
&k_{1-}^{\mu}&\Gamma^{W^-\phi^+\phi^-\phi^+}_{\mu}(k_{1-},k_{1+},k_{2-},k_{2+}) +M_W\Gamma^{\phi^-\phi^+\phi^-\phi^+}(k_{1-},k_{1+},k_{2-},k_{2+})\nonumber\\
&-&\frac{e}{2s_w}\left[\Gamma^{H\phi^-\phi^+}(k_{1+}+k_{1-},k_{2-},k_{2+})-i\Gamma^{\chi \phi^-\phi^+}(k_{1+}+k_{1-},k_{2-},k_{2+})\right.\nonumber\\
&+&\left.\Gamma^{H\phi^+\phi^-}(k_{2+}+k_{1-},k_{1+},k_{2-})-i\Gamma^{\chi \phi^+\phi^-}(k_{2+}+k_{1-},k_{1+},k_{2-})\right]=0
\eqa

\end{document}